\documentclass{iopart}
\usepackage{iopams}  
\usepackage{amsmath}
\usepackage{graphicx}
\usepackage{url}
\usepackage{booktabs}
\usepackage{multirow}
\usepackage{color}
\usepackage{hyperref}
\usepackage[normalem]{ulem}

\def\gw#1{gravitational wave#1 (GW#1)\gdef\gw{GW}}

\newcommand{\diff}{{\mathrm d}}

\newcommand{\msun}{{\mathrm M}_{\odot}}
\newcommand{\matr}[1]{\mathbf{#1}}
\newcommand{\tran}[1]{#1^{\top}}
\newcommand{\match}{{\mathcal M}}
\newcommand{\fpeak}{f_{\mathrm{peak}}}

\newcommand{\detrate}{{\dot{\mathcal N}_{\mathrm{det}}}}

\def\imbh#1{intermediate mass black hole#1(IMBH#1)\gdef\imbh{IMBH}}
\def\smbh#1{supermassive black hole#1(SMBH#1)\gdef\smbh{SMBH}}
\def\bbh#1{binary black hole#1 (BBH#1)\gdef\bbh{BBH}}
\def\bns#1{binary neutron star#1 (BNS#1)\gdef\bns{BNS}}
\def\bh#1{black hole#1 (BH#1)\gdef\bh{BH}}
\def\ns#1{neutron star#1 (NS#1)\gdef\ns{NS}}
\def\gw#1{gravitational wave#1 (GW#1)\gdef\gw{GW}}
\def\sn#1{core-collapse supernova#1 (CCSN#1)\gdef\sn{CCSN}}
\def\pnw#1{post-Newtonian#1 (PN#1)\gdef\pnw{PN}}
\def\eos#1{equation of state#1 (EoS#1)\gdef\eos{EoS}}
\def\grb#1{gamma-ray burst#1 (GRB#1)\gdef\grb{GRB}}
\def\amr#1{adaptive mesh refinement#1 (AMR#1)\gdef\amr{AMR}}
\def\isco#1{innermost stable circular orbit#1 (ISCO#1)\gdef\isco{ISCO}}
\def\cwb#1{Coherent WaveBurst#1 (CWB#1)\gdef\cwb{CWB}}
\def\snr#1{signal-to-noise ratio#1 (SNR#1)\gdef\snr{SNR}}
\def\pca#1{principal component analysis#1 (PCA#1)\gdef\pca{PCA}}
\def\mweg#1{Milky Way Equivalent Galaxy#1 (MWEG#1)\gdef\mweg{MWEG}}

\usepackage{soul} 
\usepackage{ulem} 

\begin{document}

\title{Observing Gravitational Waves From The Post-Merger Phase Of Binary
Neutron Star Coalescence}

\author{J.~A. Clark}
\address{Center for Relativistic Astrophysics and School of Physics, Georgia
Institute of Technology, Atlanta, GA 30332, USA} 
\author{A. Bauswein}
\address{Department of Physics, Aristotle University of Thessaloniki, GR-54124 Thessaloniki, Greece}
\address{Heidelberger Institut f\"ur Theoretische Studien, D-69118~Heidelberg, Germany}
\author{N. Stergioulas}
\address{Department of Physics, Aristotle University of Thessaloniki, GR-54124 Thessaloniki, Greece}
\author{D. Shoemaker}
\address{Center for Relativistic Astrophysics and School of Physics, Georgia
Institute of Technology, Atlanta, GA 30332, USA} 

\date{\today}

\begin{abstract}
    We present an  effective, low-dimensionality frequency-domain template for
    the gravitational wave signal from the stellar remnants from binary neutron
    star coalescence.  A principal component decomposition of a suite of
    numerical simulations of binary neutron star mergers is used to construct
    orthogonal basis functions for the amplitude and phase spectra of the
    waveforms for a variety of neutron star equations of state and binary mass
    configurations.  We review the phenomenology of late merger / post-merger
    gravitational wave emission in binary neutron star coalescence and
    demonstrate how an understanding of the dynamics during and after the merger
    leads to the construction of a universal spectrum.  We also provide a
    discussion of the prospects for detecting the post-merger signal in future
    gravitational wave detectors as a potential contribution to the science case
    for third generation instruments.  The template derived in our analysis
    achieves $>90\%$ match across a wide variety of merger waveforms and strain
    sensitivity spectra for current and potential gravitational wave detectors.
    A Fisher matrix analysis yields a preliminary estimate of the typical
    uncertainty in the determination of the dominant post-merger oscillation
    frequency $\fpeak$ as $\delta \fpeak \sim 50$\,Hz.  Using recently derived
    correlations between $\fpeak$ and the neutron star radii, this suggests
    potential constraints on the radius of a fiducial neutron star of $\sim
    220$\,m.  Such measurements would only be possible for nearby ($\sim
    30$\,Mpc) sources with advanced LIGO but become more feasible for planned
    upgrades to advanced LIGO and other future instruments, leading to
    constraints on the high density neutron star equation of state which are
    independent and complementary to those inferred from the pre-merger inspiral
    gravitational wave signal.  We study the ability of a selection of future
    gravitational wave instruments to provide constraints on the neutron star
    equation of state via the postmerger phase of binary neutron star mergers.
\end{abstract}

%


\section{Introduction}
The second generation of \gw{} observatories has now become operational with the
first observations by advanced LIGO (aLIGO) underway~\cite{2015CQGra..32g4001T}.
Instruments such as advanced Virgo (advVirgo)~\cite{TheVirgo:2014hva} and
KAGRA~\cite{2013PhRvD..88d3007A} will soon come online, eventually culminating
in a world-wide network of \gw{} observatories.  The \gw{} signal from the
inspiral stage of \bns{} coalescence is amongst the most promising sources for
this second generation of \gw{} detectors.  Observations of \bns{} \gw{}
inspiral signals from relatively nearby events (a few tens of Mpc) can lead to
strong constraints on the supranuclear \eos{} via the impact on the phase
evolution of the signal from tidal interactions during the latter stages of the
merger~\cite{2008PhRvD..77b1502F, 2009PhRvD..79l4033R, 2010PhRvD..81l3016H,
2010PhRvL.105z1101B, 2012PhRvD..86d4030B, 2012PhRvD..85l3007D,
2013PhRvD..88d4042R} \footnote{Reviews of the subject may also be found
in~\cite{2010CQGra..27k4002D, 2011GReGr..43..409A, 2010nure.book.....B,
lrr-2012-8, 2013rehy.book.....R}}.

For example, Read \emph{et al}~\cite{2009PhRvD..79l4033R} showed that neutron
star radii could be constrained with an uncertainty of $10\%$ for a single
nearby (100 Mpc, assuming optimal orientation and sky-location) source, based on
Fisher matrix estimates.  More recently, a number of full Bayesian analyses have
been carried out which have used astrophysically-motivated simulated populations
of \bns{} merger events to develop and probe \eos{} constraints in the low
\snr{} regime.  In~\cite{PhysRevLett.111.071101} it was found that only a few
tens of inspiral events are required to measure \ns{} tidal deformability to
$\sim 10\%$.  Similar results are found in~\cite{2015arXiv150305405A,2015PhRvD..91d3002L}, where the
\ns{} radius is determined to $\pm 1$\,km and tidal deformability is determined
to $10\mbox{-}50$\% after a few tens of \gw{} detections.

The focus of this work, however, is on the independent and complementary
constraints on the \eos{} which may be obtained from the \emph{post}-merger
signal.  Depending on the mass configuration of the system and the \eos{}, a
\bns{} merger may result in prompt collapse to a black hole (high-mass, soft
\eos{}) or the formation of a stable or quasi-stable \ns{} remnant which again,
may or may not collapse to a black hole depending on its mass and the \eos,
while transient non-axisymmetric deformations and quadrupolar oscillations in
this remnant typically give rise to a richly structured, high-frequency
(1--4\,kHz) \gw{} spectrum and a signal lasting $\sim10\mbox{-}100$\,ms,
\cite{1994PhRvD..50.6247Z, 1996A&A...311..532R, 2005PhRvL..94t1101S,
2005PhRvD..71h4021S, 2007A&A...467..395O, 2007PhRvL..99l1102O,
2008PhRvD..77b4006A, 2008PhRvD..78b4012L, PhysRevD.78.084033,
2009PhRvD..80f4037K, 2011MNRAS.418..427S, 2011PhRvD..83d4014G,
2011PhRvD..83l4008H, 2011PhRvL.107e1102S, 2012PhRvL.108a1101B,
2012PhRvD..86f3001B, 2013PhRvL.111m1101B, 2013PhRvD..88d4026H,
2013arXiv1311.4443B, 2014arXiv1403.5672T, 2000ApJ...528L..29B}.  Characterising
the frequency content of \gw{} signals from the post-merger system provides
unique opportunities for \gw{} asteroseismology: the dominant post-merger
oscillation frequency $\fpeak$ exhibits a tight correlation with the radius of
nonrotating neutron stars, with an overall uncertainty of a few hundred meters,
depending on the total binary mass. For example,  for a total binary mass of
2.7~$M_\odot$~the  uncertainty in the radius of a cold, non-rotating NS of mass
1.6\,$M_{\odot}$ (denoted as $R_{\mathrm{1.6}}$) is about
220m~\cite{2012PhRvD..86f3001B,bauswein:14}.    Similar relationships between
the dominant spectral features and stellar parameters have been confirmed
elsewhere~\cite{2013PhRvD..88d4026H,2014arXiv1403.5672T}. A deeper understanding
of the features of postmerger GW spectra has been provided
in~\cite{2011MNRAS.418..427S,bauswein:15,bauswein:july15}, where it was shown
that the spectrum is dominated by a linear feature (quadrupolar oscillations), a
quasi-linear feature (a coupling between quadrupolar and quasi-radial
oscillations) and a fully nonlinear feature (a transient spiral deformation),
leading to a classification scheme of the postmerger GW emission depending on
the EoS and binary mass.  More recently, efforts have been made to find
correlations between the pre- and post-merger signals. In
Ref.~\cite{2015arXiv150401764B} the authors derive a relation between the tidal
coupling constant $\kappa$ that determines the tidal interactions before and
during the merger and the peak frequency $\fpeak$ in the post-merger spectrum.
Thus, measurements of the inspiral signal (which determine $\kappa$) could be
used to constrain $\fpeak$ by restricting its range of possible values and by
combining measurements with those of the post-merger signal.

This connection between the tidal interactions and the post-merger oscillations
highlights the complementarity of pre- and post-merger \gw{} observations.  The
constraints arising from inspiral observations may be subject to systematic
biases induced by errors in the phase due to missing high PN-order terms or
insufficiently accurate descriptions of spin or tidal effects.  These systematic
errors can be as large as the statistical uncertainty in characterising the
inspiral signal~\cite{2014PhRvL.112j1101F,2015PhRvD..91d3002L}.  While inspiral
waveform models will continue to improve and incorporate such effects, we note
that analyses of the post-merger signal are subject to a completely independent
source of systematic error (e.g., the precise $\fpeak\mbox{-}R_{\mathrm{1.6}}$
relationship). Moreover, since the majority of pre-merger \ns{s} are likely to
have masses in the range $\sim 1.35\pm0.15$\,$\msun$, the pre-merger waveforms
are limited to probing the structure of \ns{s} in that mass-range, while the
post-merger signal allows us to probe the regime of higher masses (this is
because, e.g. the central density of the remnant of a $1.35+1.35M_{\odot}$
merger is close to the central density of a  $1.6M_{\odot}$ nonrotating star). 

This high frequency component of the merger signal, however, will be somewhat
challenging to observe in the upcoming generation of \gw{} detectors.
Typically, the most sensitive frequencies of ground-based \gw{} instruments lie
around 10--1000\,Hz, with a rapidly diminishing sensitivity in the kHz regime.
Additionally, the absence of a complete waveform model for the full pre- and
post-merger signal, or even for the post-merger signal alone, currently
prohibits the use of matched filtering and one must turn to more robust, but
ultimately less sensitive unmodelled burst searches.  For example, the study
in~\cite{2014PhRvD..90f2004C} revealed that a typical realistic burst analysis
yielded an effective range approximately 30--40\% of that which could be
possible with an optimal matched filter.

Clearly then, there is great motivation and opportunity to develop more
sensitive, more targeted analysis techniques and effective models which will
bring us closer to the sensitivity offered by a matched filtering analysis.  It
is the goal of this work to explore a \pca{} based approach to constructing
precisely such an effective model for the high-frequency component of the \bns{}
merger signal.  We construct a catalogue of 50 numerical waveforms from the
merger and post-merger evolution of a variety of \bns{} systems with various
\eos{s} and mass configurations.  The magnitude and phase spectra of the
waveforms in the catalogue are then decomposed into orthogonal bases using
\pca{}.  These basis functions can then be used to construct a frequency-domain
waveform template which provides, on average, a 93\% match for the waveforms in
the catalogue for both aLIGO and a variety of potential upgrades and new
\gw{} instruments.

The structure of this paper is as follows: in section~\ref{sec:properties} we
provide a detailed review of \bns{} merger and post-merger phenomenology,
focussing on the resulting features in the \gw{} spectrum and hence how one may
constrain the \ns{} \eos{}.  Section~\ref{sec:detectability} summarises the
expected detectability of the high-frequency \bns{} waveforms used in this
study, assuming a matched-filtering approach and a variety of current and
potential future \gw{} instruments.  In section~\ref{sec:pca} we describe and
characterise our \pca{}-based frequency-domain waveform template in terms of
waveform match and provide Fisher-matrix estimates of the uncertainties in
$\fpeak$ and $R_{\mathrm{1.6}}$ based on this template as a \emph{preliminary}
guide to its potential.  Finally, section~\ref{sec:conclusion} provides a
summary and some concluding remarks relating to the planned applications of this
model and the potential for similar approaches to enhance unmodelled burst
analyses.

\section{Properties of Postmerger GW Spectra and Constraining the Neutron Star EOS}
\label{sec:properties}

\subsection{Types of merger dynamics and GW spectra}

\begin{figure}
    \includegraphics[width=1\columnwidth]{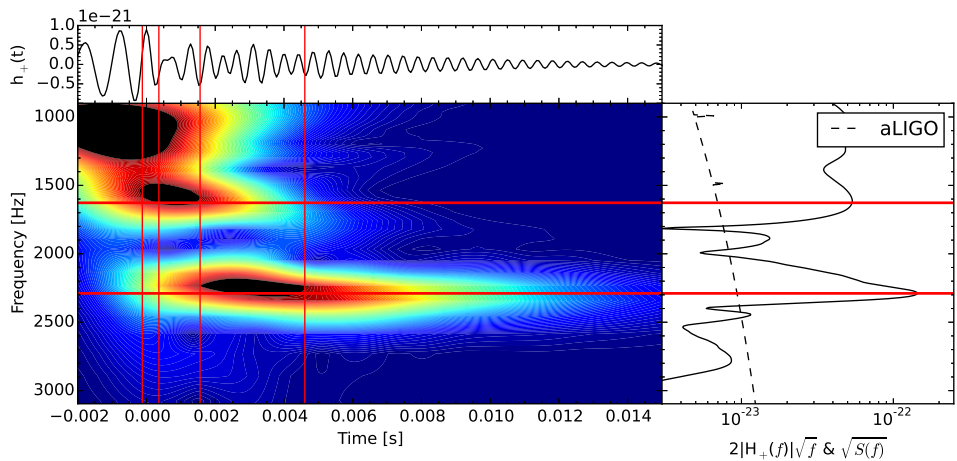}
    \caption{\label{fig:timefreq_tm1}Time-frequency analysis for the  TM1 1.35+1.35
        waveform for an optimally-oriented source at 50\,Mpc.  The top and right
        panels show the time-domain waveform-component $h_+$  and its Fourier
        magnitude spectrum, respectively.  The time-frequency map is constructed
        from the magnitudes of the coefficients of a continuous wavelet
        transform using a Morlet basis.  Horizontal red lines emphasise the
        locations of the peak frequency $\fpeak$ and the secondary peak which,
    in this case, corresponds to $f_{\mathrm{spiral}}$ (see text and
figure~\ref{fig:snap}). The vertical lines correspond to the time steps of the
four panels in Fig.~\ref{fig:snap}. }
\end{figure}

For symmetric (i.e., equal component masses) and mildly asymmetric binaries the
\gw{} postmerger spectra of NS mergers (see e.g. right panel of
Fig.~\ref{fig:timefreq_tm1} or Fig.~1 in Ref.~\cite{bauswein:15}) show a generic behavior
in the sense that certain features of the spectrum depend in a particular way on
the total binary mass and the high-density EoS~\cite{bauswein:15}. Specifically,
distinct peaks in the spectrum can be associated with distinct mechanisms
generating those features, and the frequency and strength of the different \gw{}
peaks are determined by the total binary mass and \eos{}.  The presence or absence
of certain secondary peaks in the spectrum, together with their relative
strengths is determined by the quasi-linear coupling between the quasi-radial
and quadrupolar oscillation modes and by the orbital motion of antipodal bulges
of a spiral deformation in the remnant.    The characteristics of these distinct
spectral features can be used to classify the post-merger dynamics of the
system~\cite{bauswein:15}.

The most striking feature of the postmerger spectrum is a major peak generated
by the dominant quadrupolar oscillation of the remnant, which is present in all
models that form a NS merger remnant. The determination of the frequency of this
peak in a GW measurement is the focus of this work because the peak frequency
scales tightly with the radii of nonrotating \ns{s} (see Fig.~\ref{fig:freqrel}
below and discussion in~\cite{2012PhRvL.108a1101B,bauswein:12}) and thus
provides strong constraints on the only incompletely known EoS of NS matter.
Apart from this main peak, there can be up to two pronounced secondary peaks at
frequencies below $f_\mathrm{peak}$.

\begin{figure}
    \includegraphics[width=0.4\columnwidth,angle=270]{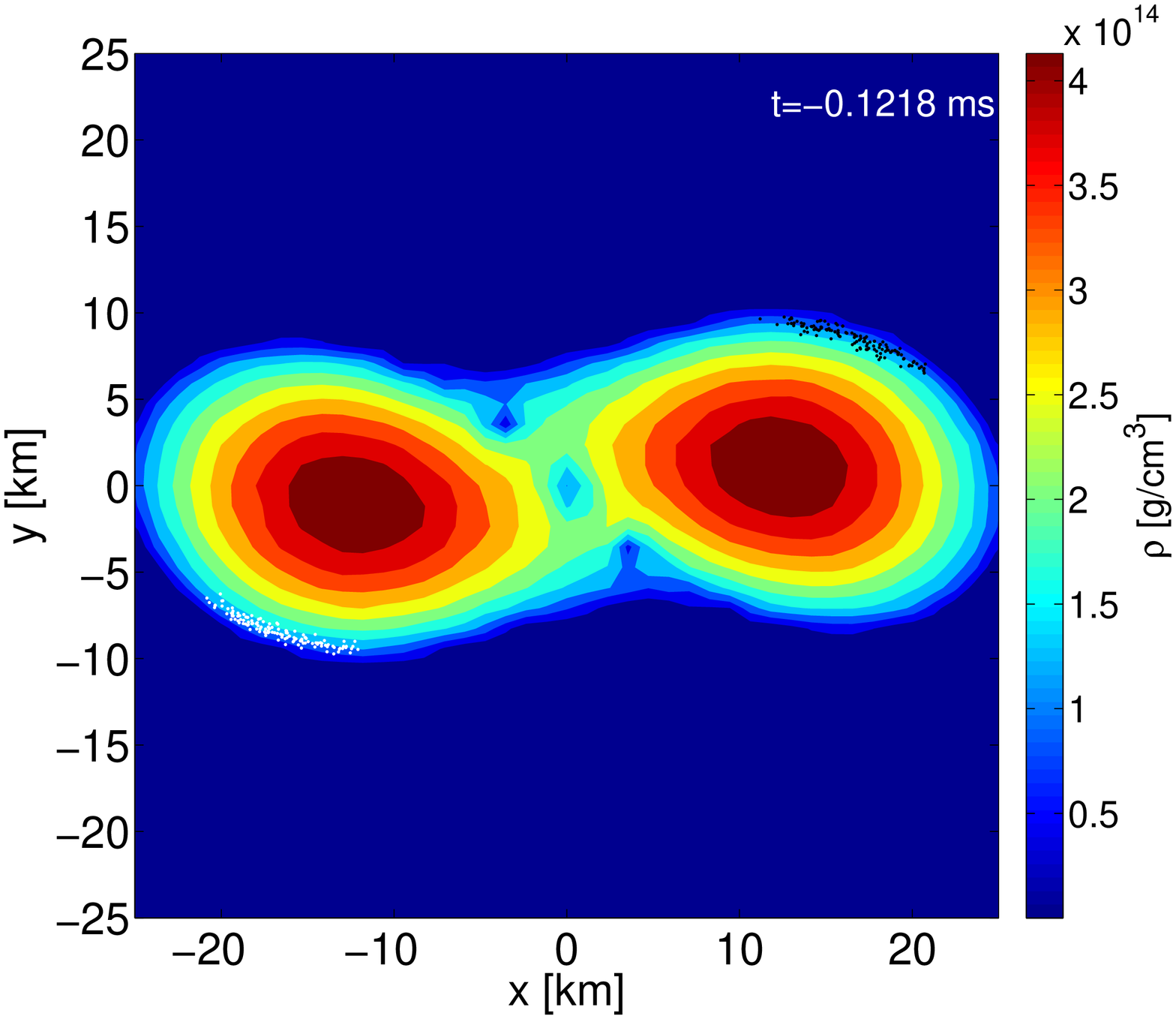}
    \includegraphics[width=0.4\columnwidth,angle=270]{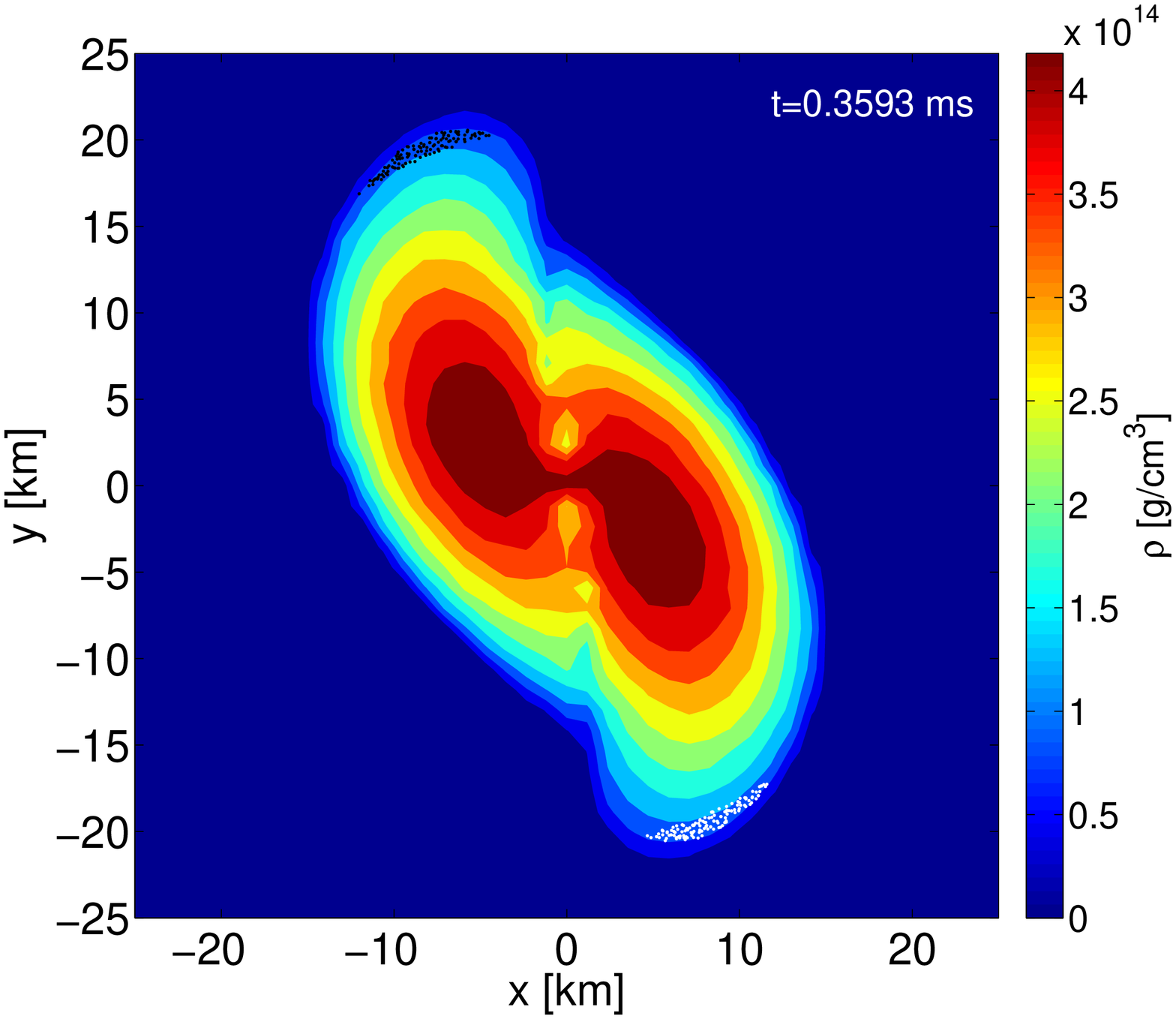}\\
    \includegraphics[width=0.4\columnwidth,angle=270]{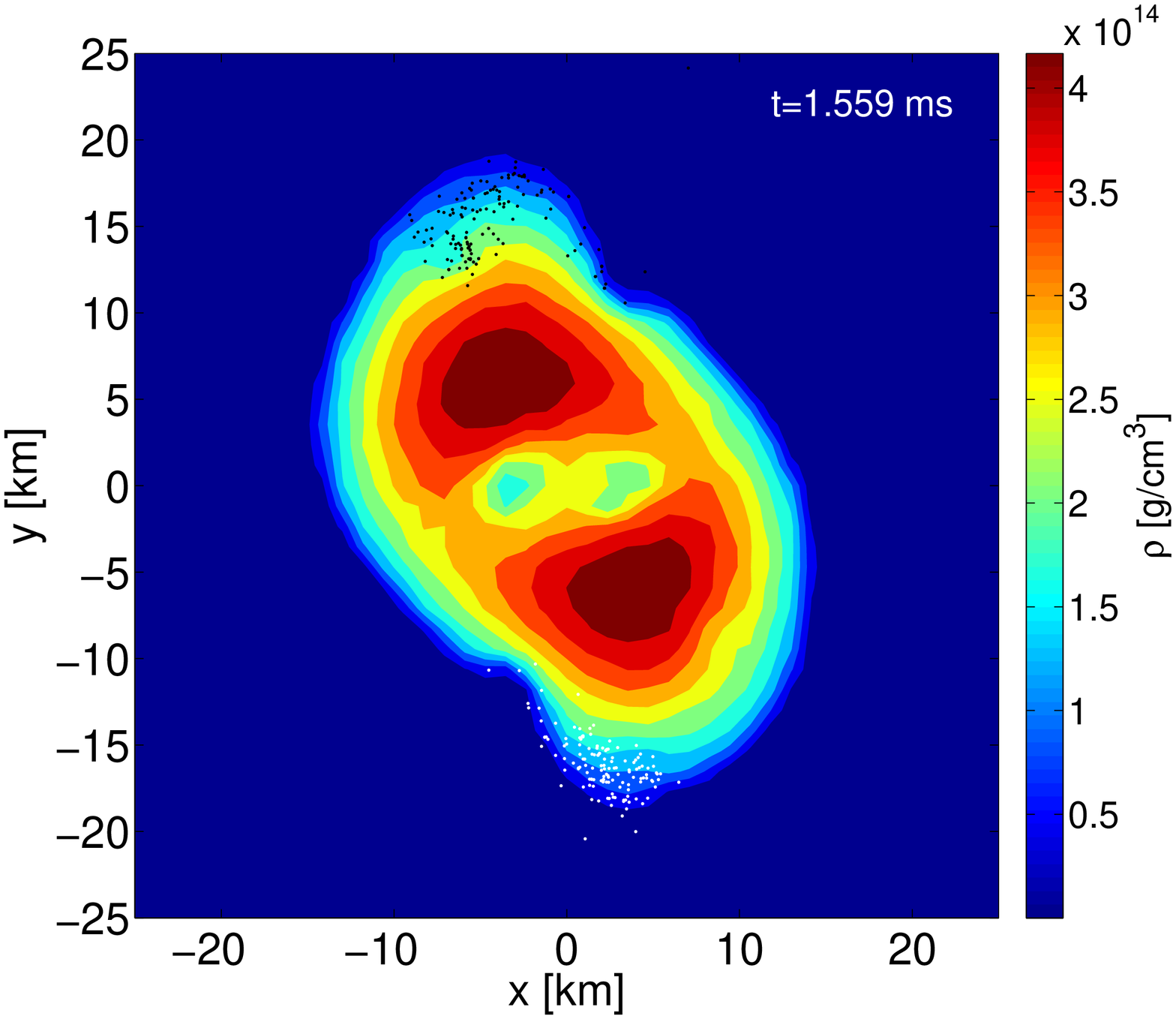}
    \includegraphics[width=0.4\columnwidth,angle=270]{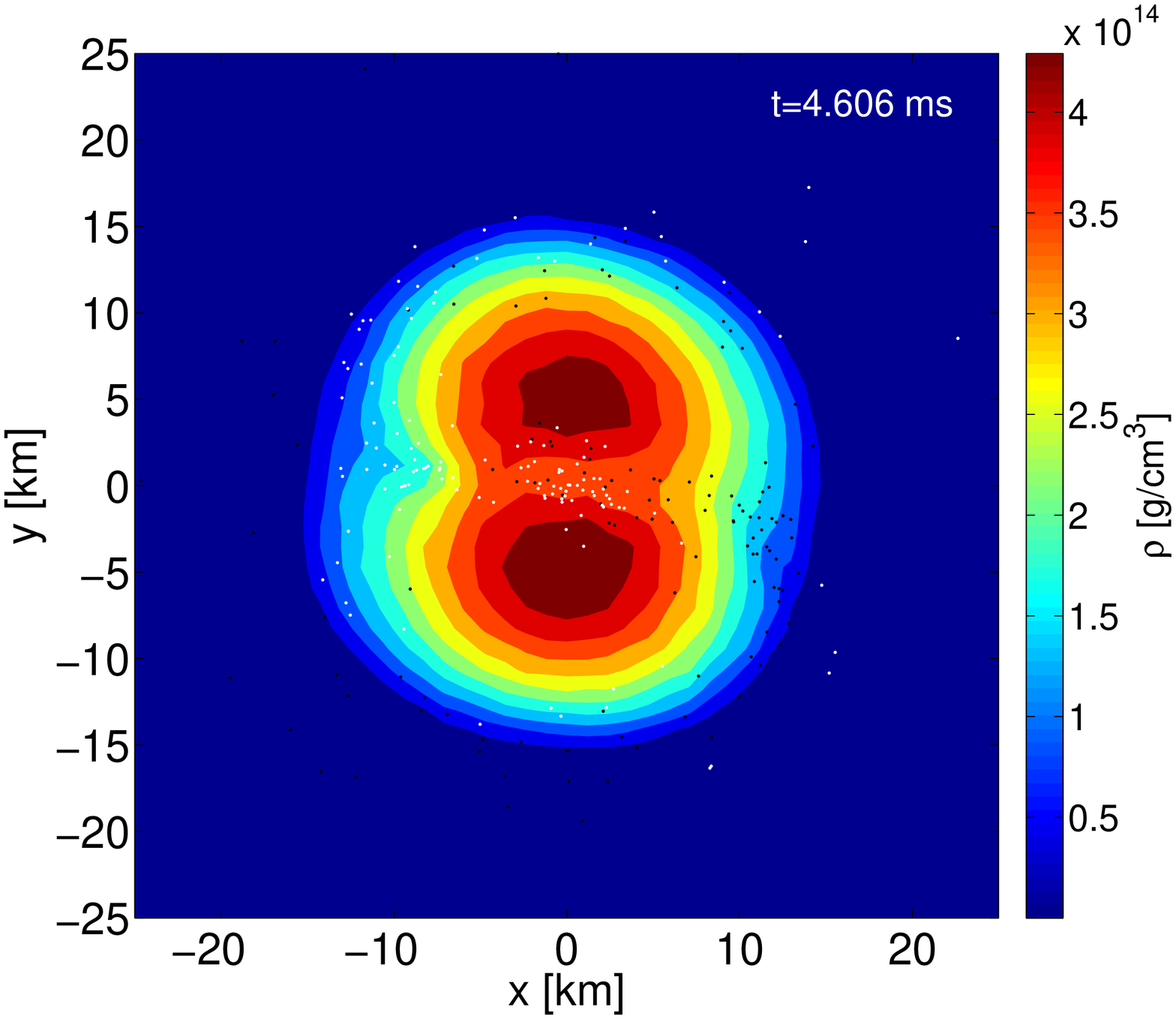}  
    \caption{\label{fig:snap}Evolution of the rest-mass density in the
    equatorial plane for a 1.35-1.35~$M_\odot$ merger with the TM1
EoS~\cite{1994NuPhA.579..557S,2012ApJ...748...70H}. Black and white dots
indicate the positions of selected fluid elements constituting the antipodal
bulges, which generate a distinct peak in the GW spectrum. (A low number of
iso-density contours is chosen for a better identification of the different
remnant components. This choice leads to an artificially coarse visualization of
the simulation data.)}
\end{figure}

One of the secondary features is a peak generated by the quasi-linear
interaction between the dominant quadrupolar oscillation and the quasi-radial
mode of the remnant (the latter does not appear strongly in the GW spectrum on
its own)~\cite{2011MNRAS.418..427S}. The corresponding peak of the quasi-linear
mode coupling has an amplitude proportional to the product of the amplitudes of
the quadrupolar mode and of the quasi-radial mode, while its frequency, which we
denote as $f_{2-0}$, is equal to the difference of the frequencies of these two
modes, i.e.  $f_{2-0}=f_\mathrm{peak}-f_0$, where $f_0$ is the frequency of the
quasi-radial mode. The $f_{2-0}$ feature is particularly pronounced for
relatively high total binary masses and soft \eos{s}. Another secondary spectral
peak is produced by the orbital motion of antipodal bulges, which form during
the merging as a spiral deformation and then orbit around the inner remnant for
a few milliseconds~\cite{bauswein:15} (see Fig.~\ref{fig:snap}).  This dynamical
feature is present in addition to the main emission at $\fpeak$ for the first
few milliseconds after merging. Bulges moving with an orbital frequency
$f_\mathrm{bulges}$ result in a peak in the GW spectrum at $f_\mathrm{spiral}=2
f_\mathrm{bulges}$. This finding receives further support by the time-frequency
map of the GW signal shown in Fig.~\ref{fig:timefreq_tm1} for the
1.35-1.35~$M_\odot$ merger with the TM1
EoS~\cite{1994NuPhA.579..557S,2012ApJ...748...70H}. One can clearly recognize
that in the early postmerger phase there are two distinct frequencies
simultaneously contributing to the GW signal. The frequency of the dominant
remnant oscillation is present for many milliseconds. The secondary peak at
$f_\mathrm{spiral}$ is generated within the first few milliseconds, when the
antipodal bulges are pronounced (see Fig.~\ref{fig:snap}). There is no evidence
for a strong time variation of the frequencies, especially of the dominant
frequency, which was suggested as an explanation for the structure of the GW
spectrum in~\cite{2014arXiv1411.7975K,2014arXiv1412.3240T}.

\begin{figure}
    \includegraphics[width=1\columnwidth]{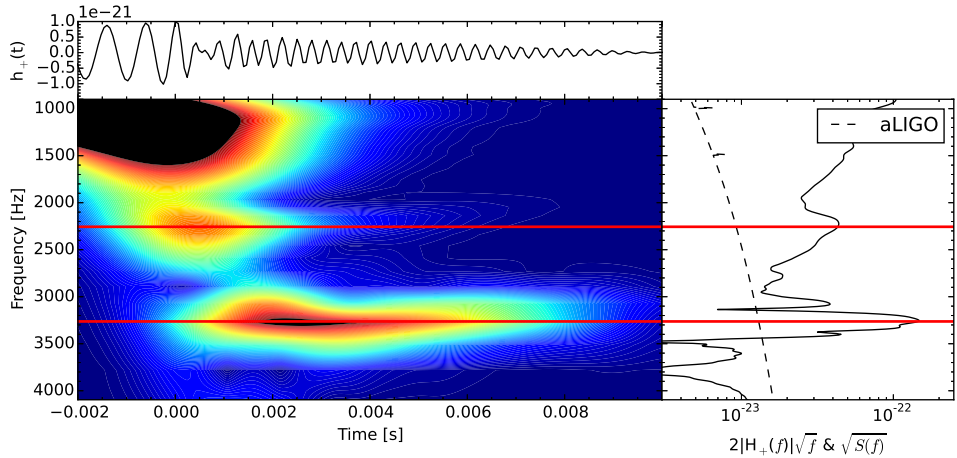}
    \caption{\label{fig:timefreq_sfho}Time-frequency analysis for the  SFHO 1.35+1.35
        waveform for an optimally-oriented source at 50\,Mpc.  In contrast to
    figure~\ref{fig:timefreq_tm1}, the secondary spectral feature now arises
mostly from the $f_{2-0}$ oscillation, rather than the $f_{\mathrm{spiral}}$ feature
from the antipodal bulges.}
\end{figure}

The information in the time-frequency map of the GW signal can be related to the
dynamical behavior of the remnant, which we illustrate by the evolution of the
rest-mass density in the equatorial plane for the same simulation (see
Fig.~\ref{fig:snap}). The time step of the different snapshots are marked in the
time-frequency map (Fig.~\ref{fig:timefreq_tm1}) by vertical lines.  Evidently, the
presence of antipodal bulges at the outer remnant coincides with the presence of
power at $f_\mathrm{spiral}$ in the time-frequency map.  It is apparent that the
$f_\mathrm{spiral}$ feature is initially particularly strong exceeding even the
emission at $f_\mathrm{peak}$ ; the antipodal bulges are strongest during and
immediately after merging and  the spiral deformation forming the bulges
initially comprises large parts of the remnant (see upper right panel in
Fig.~\ref{fig:snap}).  In Fig.~\ref{fig:snap}, the antipodal bulges complete
approximately one orbit from the top right to the bottom left panel in about
1.2~ms. Thus, the orbital frequency
$f_\mathrm{bulges}=1/1.2~\mathrm{ms}=0.833$~kHz is expected to produce a peak at
$f_\mathrm{spiral}=2 f_\mathrm{bulges}=1.67$~kHz, where a peak is found in the
spectrum (see Fig.~\ref{fig:timefreq_tm1}).  For comparison we
    also show the time-frequency analysis for the SFHO \eos{} and component
    masses 1.35-1.35$\msun$ in figure~\ref{fig:timefreq_sfho}. Here the secondary
    peak at 2.2~kHz likely arises from the $f_{2-0}$ feature. An examination of the hydrodynamical 
    data for this model reveals an $f_\mathrm{bulges}$ of about 1.25~kHz (resulting 
    in $f_\mathrm{spiral}\approx 2.5$~kHz), whereas the frequency of the quasi-radial mode 
    is $f_0=1.0$~kHz, and thus the $f_{2-0}$ peak is expected to occur at about 2.2~kHz.

The above findings on the time-frequency characteristics of the
$f_\mathrm{spiral}$ peak are consistent with the explanations of its origin
presented in \cite{bauswein:15}. The  $f_\mathrm{spiral}$ feature is
particularly strong for mergers with relatively low binary masses and stiff
\eos{s} because less compact \ns{s} favor the spiral deformation and the formation
of the antipodal bulges during merging.  In contrast, binaries with more massive
components, i.e. very compact stars, merge with a higher impact velocity, which
favours a strong excitation of the quasi-radial mode of the remnant, leading to
a strong $f_{2-0}$ feature, while the spiral deformation becomes less
pronounced.  For intermediate cases, i.e.  moderately high binary masses, both
secondary peaks are clearly present with comparable strength and distinguishable
in frequency.  Overall, this implies that for a given \eos{} the binary mass
determines the presence and strength of the different secondary features.
According to the classification scheme introduced in \cite{bauswein:15}, one can
identify three different types of spectra: High-mass/soft EOS binaries produce
spectra where the dominant secondary peak is  $f_{2-0}$ (Type I mergers). For
intermediate binary masses and EOS stiffness, both the $f_\mathrm{spiral}$ and
$f_{2-0}$ features are present with roughly comparable amplitude (Type II
mergers).  Low-mass/stiff EOS binaries produce spectra with a strong
$f_\mathrm{spiral}$ peak and an absent $f_{2-0}$ feature (Type III mergers).
See~\cite{bauswein:15,bauswein:july15} for further discussion.

\subsection{Universal Post-merger Spectra \& Measuring The Neutron Star Radius}
\label{sec:spectra2radii}

\begin{figure}
    \includegraphics[width=0.48\columnwidth]{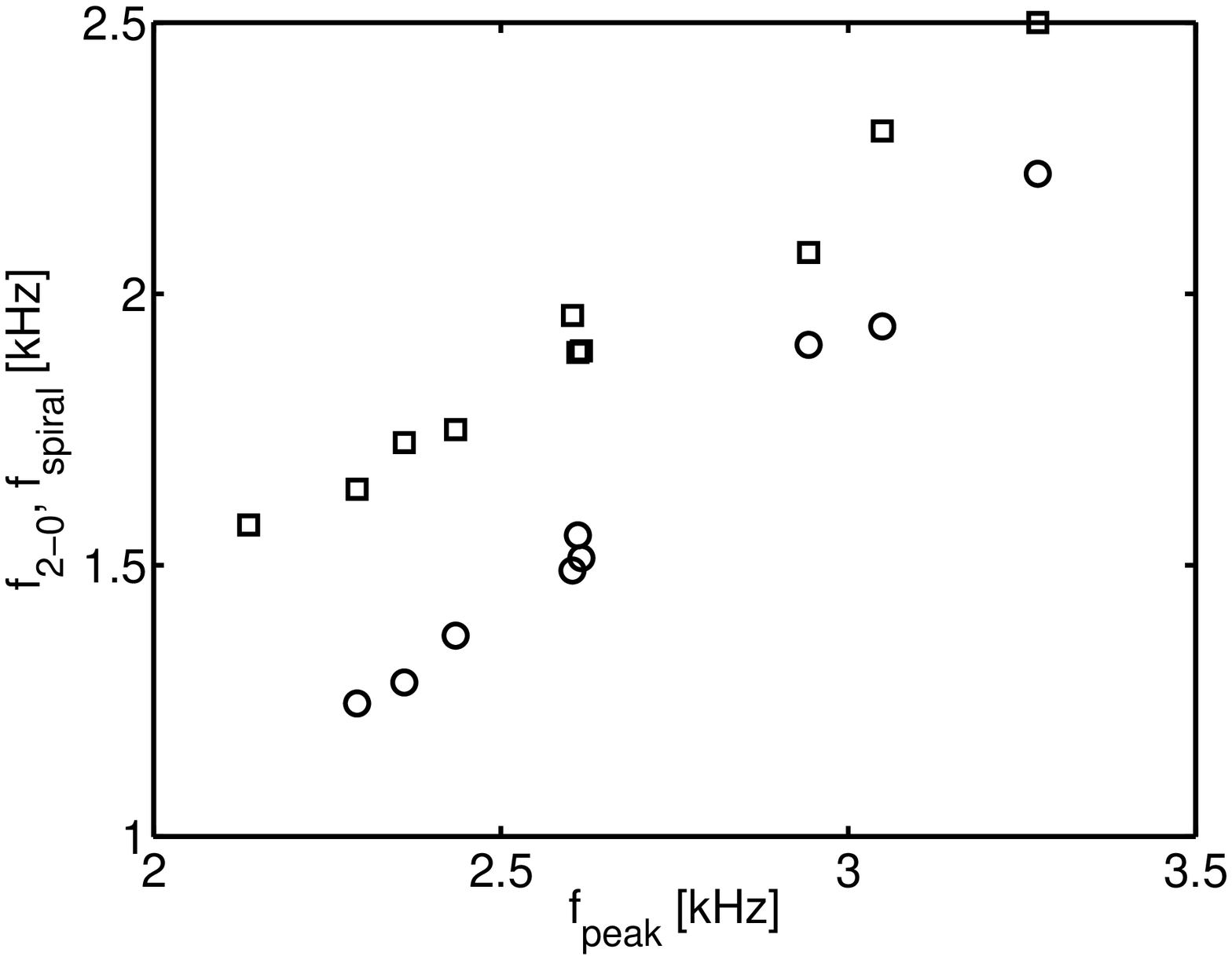}
    \includegraphics[width=0.48\columnwidth]{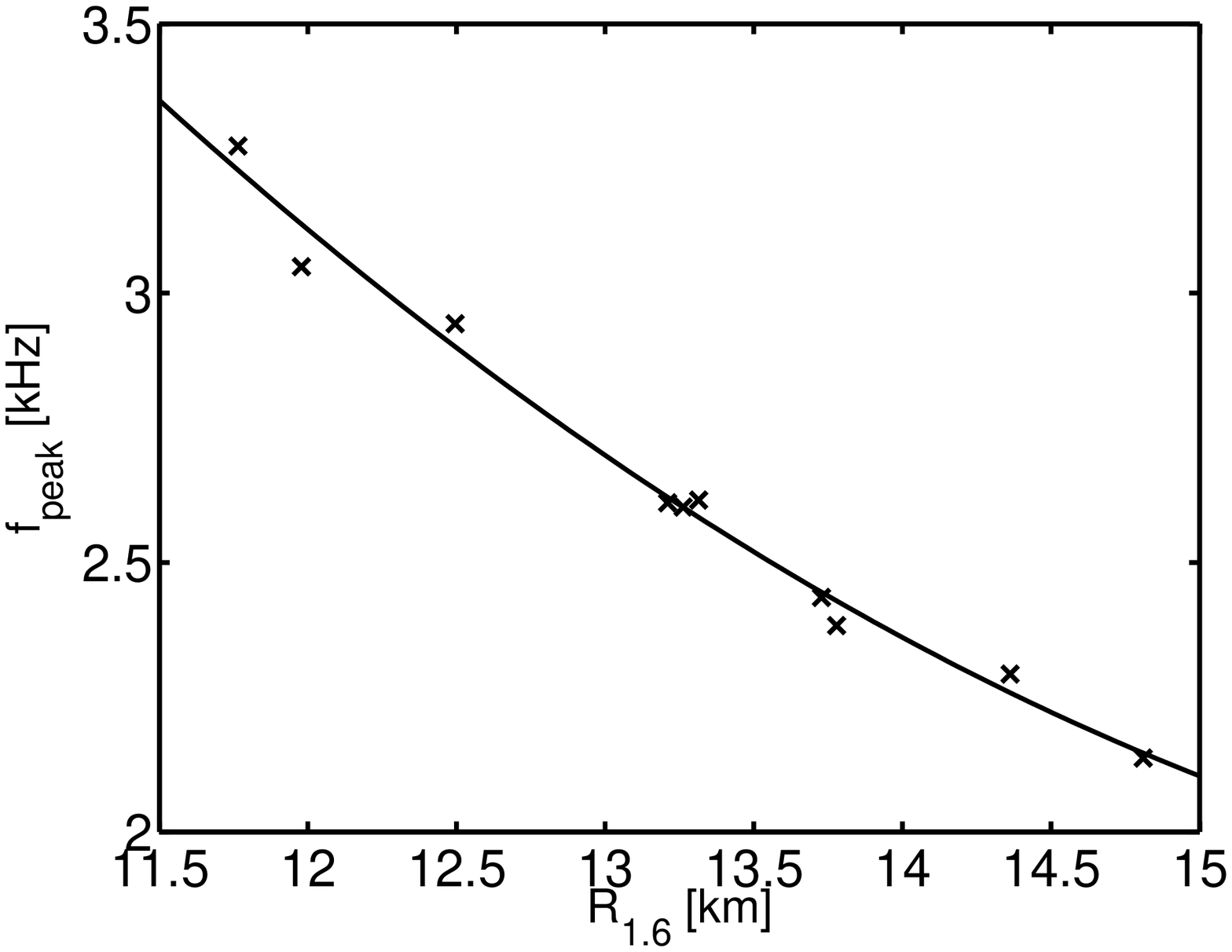}
    \caption{\label{fig:freqrel} Left panel: Secondary frequencies $f_{2-0}$
        (circles) and $f_\mathrm{spiral}$ (boxes) and as function of the
        dominant postmerger oscillation frequency $f_\mathrm{peak}$ for
        1.35-1.35~$M_\odot$ mergers with different EoSs (reproduced
        from~\cite{bauswein:15}). Right panel: Peak frequency as a function of
        radii of nonrotating NSs with 1.6~$M_{\odot}$ for 1.35-1.35~$M_\odot$ mergers
        with different EoSs. Solid line shows a least-square fit to the data.}
\end{figure}

For a fixed total binary mass the frequencies of the three different peaks
depend in a particular way on the EoS, which can be characterized by the radius
or compactness of nonrotating
\ns{s}~\cite{2012PhRvL.108a1101B,bauswein:12,bauswein:15} (see
also~\cite{2014PhRvL.113i1104T,2014arXiv1412.3240T} for the dependence of the
strongest secondary feature on compactness, without distinguishing the different
nature of secondary peaks). The two secondary frequencies show a tight
correlation with the dominant postmerger frequency $f_{peak}$. This is shown in
Fig.~\ref{fig:freqrel} for 1.35-1.35~$M_\odot$ mergers.

\begin{figure}
 \includegraphics[width=0.48\columnwidth]{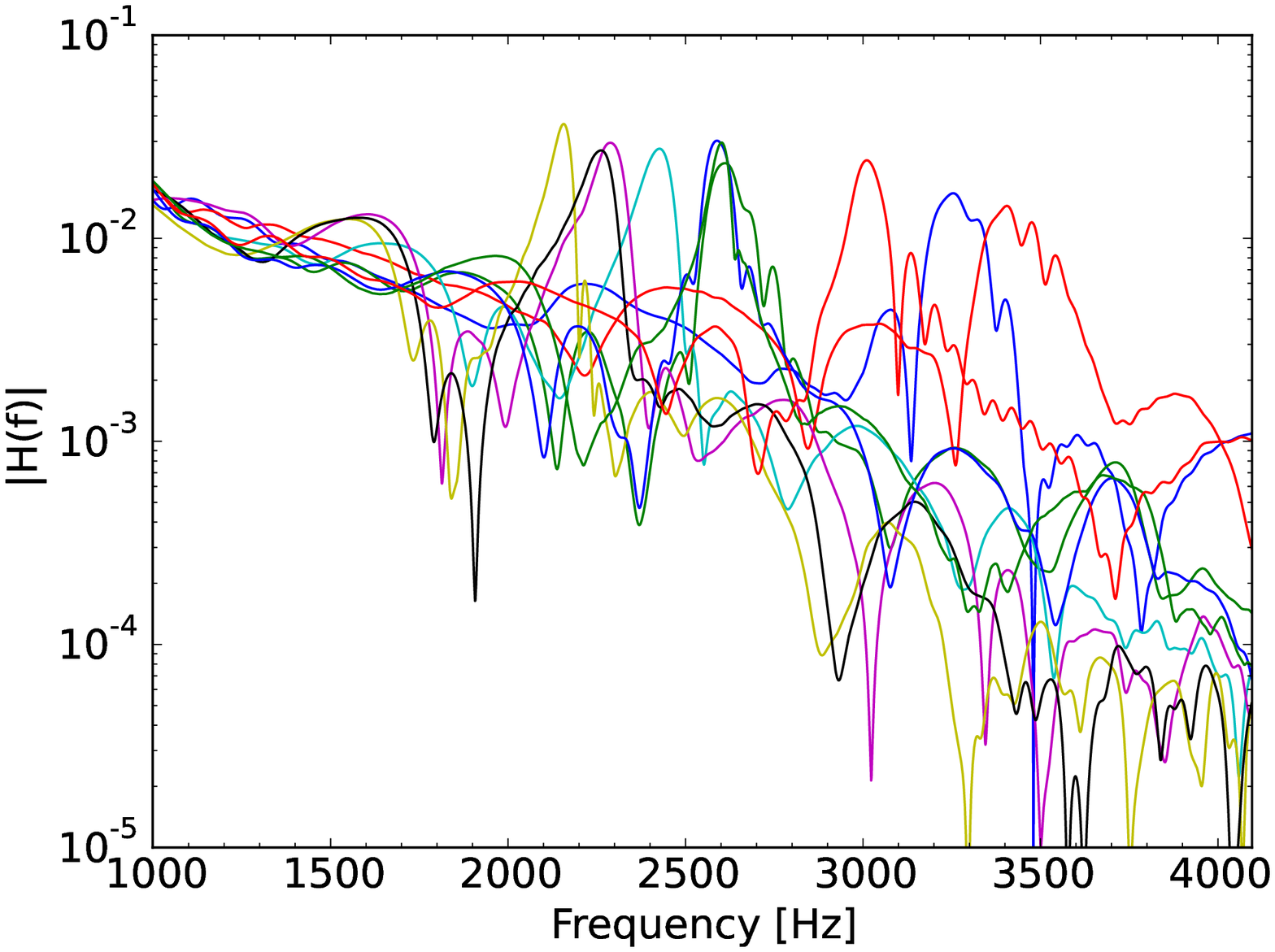}
 \includegraphics[width=0.48\columnwidth]{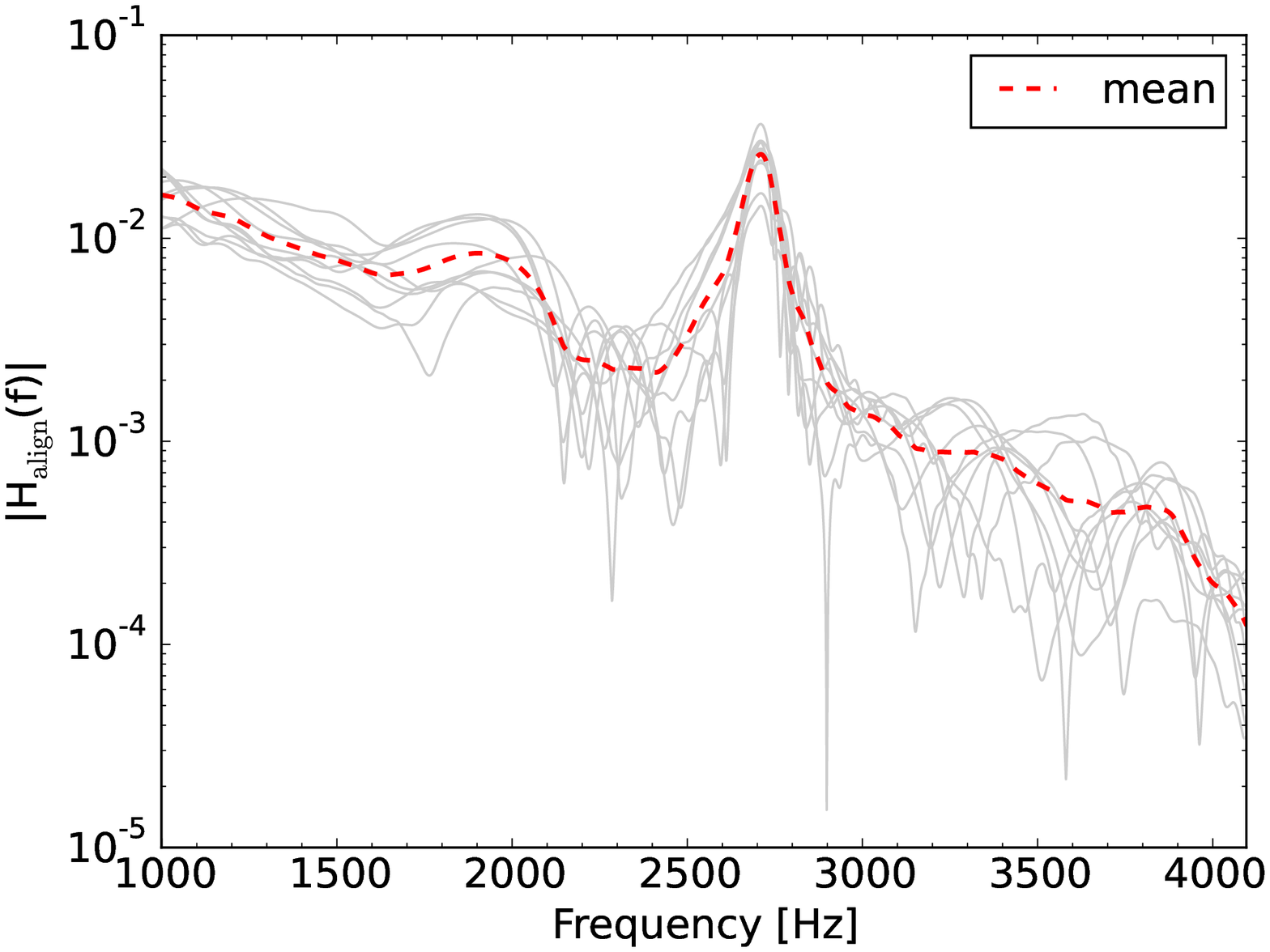}
 \caption{\label{fig:universal_spectra}\emph{Left}: \gw{} spectra from
     1.35-1.35\,$M_\odot$ \bns{} mergers with various \eos{s}.  Waveforms have
     been normalised to unit $h_{\mathrm{rss}}$ (see equation~\ref{eq:hrss}).
     \emph{Right}: The same spectra, now rescaled, using the procedure described
 in detail in \S~\ref{sec:pca}, such that their peak frequencies are aligned.
 Note that this also quite effectively aligns the secondary features.}
\end{figure}

The existence of generic spectral features with predictable behavior suggests
that the construction of a universal spectrum should be feasible through the
appropriate alignment of the main peaks from spectra for various \eos{s}.
Figure~\ref{fig:universal_spectra} shows the \gw{} spectra for equal mass
binaries (1.35-1.35$\msun$) with different \eos{s}.  The waveforms have been
normalised such that the root-sum-squared amplitude is unity,
\begin{equation}\label{eq:hrss}
    h_{\mathrm{rss}} = \int_{-\infty}^{\infty} |h(t)|^2\diff t = 1
\end{equation}
In the right panel, we
rescale the frequency axis such that the dominant quadrupolar oscillation peak
feature is located at a common reference value (2.6\,kHz) for all models.  
Apart from small variations of the secondary features a remarkable
universality of the spectra is found, which can be explained as follows: We
choose a reference peak frequency of $f_{\mathrm{ref}}=2.6$~kHz. Thus, for a
spectrum with the main peak at $f_{\mathrm{peak}}$ the factor for rescaling the
frequency is $a=f_{\mathrm{ref}}/f_{\mathrm{peak}}$. This factor $a$ is also
applied to the frequencies of the secondary peaks. Therefore, a rescaled
secondary peak $f_{\mathrm{sec}}$ (i.e. $f_{\mathrm{spiral}}$ or $f_{2-0}$) is
located at $a f_{\mathrm{sec}}=f_{\mathrm{ref}}
f_{\mathrm{sec}}/f_{\mathrm{peak}}$.  Since the fraction
$f_{\mathrm{sec}}/f_{\mathrm{peak}}\equiv c$ is approximately constant and
similar for both secondary features (see Fig.~\ref{fig:freqrel}), a rescaled
secondary feature occurs at approximately the same frequency $c \cdot
f_{\mathrm{ref}}$ for all models.

This universality of the scaled spectra suggests that it should be possible to
produce a model from the mean spectrum, computed over a number of numerical
simulations, plus some small deviations.  In Sec.~\ref{sec:pca}, we demonstrate
that \pca{} provides an approach to solve exactly this problem by producing an
orthornomal basis constructed from a superposition of the mean-centered spectra.
Furthermore, we find that the perturbations from the mean spectrum are generally
well described by a small number of basis functions.

It is important to stress here that, for distances which allow for the detection
of postmerger \gw{} emission, the individual masses of the binary components can
be determined with an accuracy of a few per
cent~\cite{2013ApJ...766L..14H,2014ApJ...784..119R,2015arXiv150805336F}.
Furthermore, current observations suggest that \bns{} mass configurations will not
be dramatically asymmetric (see e.g. the compilation of \ns{} masses
in~\cite{2012ARNPS..62..485L}).  Peak frequencies which are recovered within
this data analysis study, are converted to NS radii via
\begin{equation}\label{eq:freqrel}
    R_{1.6} = a \fpeak^2 + b \fpeak + c,
\end{equation}
describing the empirical relation between NS radii and the dominant postmerger
oscillation frequency for symmetric mergers for total binary masses of
2.7~$M_\odot$\footnote{We note that, even in the
abseence of a measurement of mass ratio, the $\fpeak\mbox{-}R_{1.6}$ relation is
quite robust for a constant chirp mass, which is generally recovered to high
precision.  See e.g.,~\cite{bauswein:july15}}. Here we adopt the coefficients $a=1.099$, $b=-8.574$ and
$c=28.07$ from previous work~\cite{bauswein:14}.  For this particular total
binary mass the maximum uncertainty in the empirical relation is
175\,m.  In addition to this systematic error, the measurement of $\fpeak$ in
noisy \gw{} data will introduce a statistical error, whose determination is one
of the major goals of this work and is quantified in Sect.~\ref{sec:pca}.

\subsection{Binary Neutron Star Merger Waveforms Used In This
Study}\label{sec:waveforms} 
\begin{figure}
    \includegraphics[width=1\columnwidth]{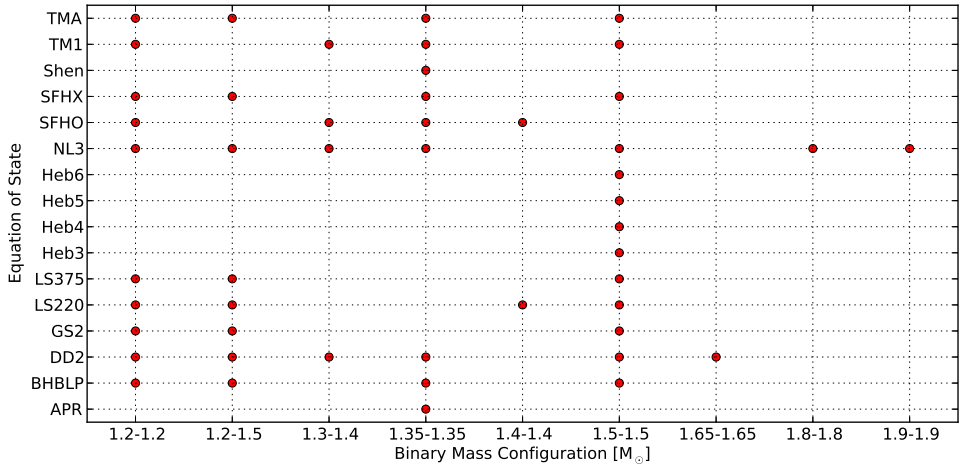}
    \caption{\label{fig:catalogue}\eos{}--mass configuration parameter space for
    the \bns{} merger waveforms used in this study.  No specific criteria were
applied in selecting waveforms, apart from a desire to cover a reasonable
variety of waveform morphologies and values for $\fpeak$.  See Table~\ref{tab1} for references of the EoSs used.}
\end{figure}

The numerical waveforms used in this study rely
mostly on the calculations discussed
in~\cite{2012PhRvL.108a1101B,bauswein:12,bauswein:14,2014PhRvD..90f2004C,bauswein:15},
where further information can be found. Additional waveform models employed here
are obtained within the same physical and numerical model, for which further
details are provided
in~\cite{2002PhRvD..65j3005O,2007A&A...467..395O,2010PhRvD..82h4043B,2012PhRvD..86f3001B}.
The \eos{} models for the hydrodynamical simulations are chosen to cover a large
variety including very stiff and very soft \eos{s} (see Table~\ref{tab1}) and a variety of binary mass
configurations are used. All \eos{s} are compatible with a maximum \ns{} mass of
$\sim 2~M_\odot$~\cite{2010Natur.467.1081D,Antoniadis26042013}.
Figure~\ref{fig:catalogue} illustrates the selection of waveforms used in this
study in terms of their \eos{s} and mass configuration.

\begin{table}
    \centering
 \begin{tabular}{llc}
\hline    
EoS & Ref. & $R_{1.35}$ [km]\\ 
\hline    
NL3   & \cite{1997PhRvC..55..540L,2010NuPhA.837..210H}   &  14.75 \\
LS375 & \cite{1991NuPhA.535..331L}                       &  13.65 \\
DD2   & \cite{2010PhRvC..81a5803T,2010NuPhA.837..210H}   &  13.21 \\
TM1   & \cite{1994NuPhA.579..557S,2012ApJ...748...70H}   &  14.49 \\
SFHX  & \cite{2012arXiv1207.2184S}                       &  11.98 \\
GS2   & \cite{2011arXiv1103.5174S}                       &  13.38 \\
SFHO  & \cite{2012arXiv1207.2184S}                       &  11.92 \\
LS220 & \cite{1991NuPhA.535..331L}                       &  12.73 \\
TMA   & \cite{1995NuPhA.588..357T,2012ApJ...748...70H}   &  13.86 \\
APR   & \cite{1998PhRvC..58.1804A}                       &  11.33 \\
BHBLP & \cite{2014ApJS..214...22B}                       &  13.21 \\
Shen  & \cite{1998NuPhA.637..435S}                       &  14.64 \\
Heb6  & \cite{2010PhRvC..82a4314H}                       &  13.33 \\
Heb5  & \cite{2010PhRvC..82a4314H}                       &  12.38 \\
Heb4  & \cite{2010PhRvC..82a4314H}                       &  12.51 \\
Heb3  & \cite{2010PhRvC..82a4314H}                       &  12.03 \\
\hline
\end{tabular}
\caption{\label{tab1} References for the EoSs used in this study. $R_{1.35}$ is
the circumferential radius of a non-rotating NS with a gravitational mass of
1.35~$M_\odot$.}
\end{table}

\section{Detectability}
\label{sec:detectability}
We now discuss the expected detectability of the post-merger \gw{} signal in
current and planned \gw{} instruments.  A natural, preliminary, measure of
detectability is the matched-filter \snr{} one would obtain given a perfect
model, or template, for the signal waveform $h$ in \gw{} detector data $s$.  The
matched-filter \snr{} is defined as,
\begin{equation}
    \rho = \frac{(s|h)}{\sqrt{(h|h)}}.
\end{equation}
The \emph{optimal} \snr{}, where the template $h$ exactly matches detector
output is then simply,
\begin{equation}
    \rho_{\mathrm{opt}} = \sqrt{(h|h)},
\end{equation}
where $(.|.)$ is the usual inner product~\cite{1994PhRvD..49.2658C}:
\begin{equation}\label{eq:inner_product}
    (a|b) = 4 \mathrm{Re} \int_{f_{\mathrm{low}}}^{f_{\mathrm{Ny}}}
        \frac{\tilde{a}(f)\tilde{b}^*(f)}{S_h(f)}~\diff f,
\end{equation}
and $S_h(f)$ is the noise spectrum of a given \gw{} detector and the asterix
indicates complex conjugation.  Note that we impose a lower bound
$f_{\mathrm{low}}$ on the frequency over which the inner product is evaluated in
order to target the detectability of the high-frequency part of the signal.  In
this study we use $f_{\mathrm{low}}=1$\,kHz.  The inner product is evaluated up
to the Nyquist frequency of the spectrum, 8192,Hz in this study.  We also
characterise detectability in terms of \emph{horizon distance}
$D_{\mathrm{hor}}$: the distance at which an optimally oriented source yields an
\snr{} at least as large as some nominal threshold, $\rho_*$.  For \gw{}
searches in which the time of arrival of the signal and the source sky-location
are unknown, it is typical to evaluate horizon distances with $\rho_*=8$.  In
our application, however, we envisage a hierachical `triggered' analysis,
similar to that described in~\cite{2014PhRvD..90f2004C}, wherein the earlier,
lower-frequency inspiral portion of the coalescence signal has already been
detected at high confidence.  It is likely then that the time of coalescence has
been determined to an accuracy of a few or a few 10's of milliseconds and we can
significantly reduce the threshold used to define the horizon distance.
Following~\cite{2014PhRvD..90f2004C}, we choose $\rho_*=5$.  Finally, we can
determine the rate $\detrate$ with which we will obtain signals with $\rho_*\geq
5$ from the expected number of \bns{} mergers which are accessible to a search
with a given horizon distance~\cite{2010CQGra..27q3001A}.  For the purposes of
this study, we assume the `realistic' rate of \bns{} coalescence
from~\cite{2010CQGra..27q3001A}:
$\mathcal{R}_{\mathrm{re}}=100$\,MWEG$^{-1}$Myr$^{-1}$.

We now compute each of the figures of merit (the \snr{} for an optimally
oriented source at 50\,Mpc; the horizon distance assuming an \snr{} threshold
$\rho_*=5$ and the expected detection rate $\detrate$) for
aLIGO~\cite{aLIGOcurves,2015CQGra..32g4001T}, as well as the following selection
of proposed upgrades to aLIGO and new facilities.  The following descriptions
emphasise the expected increases in sensitivity relative to aLIGO only over
1--4\,kHz; the band of interest for the post-merger signal.  Comparisons with
the increased range and sensitivity to the earlier inspiral part of the signal
are left to future studies.   Note also that we take aLIGO to be the most
sensitive of the second generation \gw{} detectors; instruments such as advVirgo
and Kagra offer comparable or reduced sensitivity in the frequency regime of
interest to this study.  It should be noted, however, that a \emph{network} of
$X$ detectors with comparable sensitivity could improve the range of an search
by a factor of up to $\sim\sqrt{X}$ with respect to the single detector
expectation, assuming stationary Gaussian noise and an optimal analysis.  We
restrict our estimates to single detector ranges and rates in the interests of
conservatism and simplicity.
\begin{description}
    \item [LIGO A+~\cite{2015PhRvD..91f2005M,ISwhitePaper}] 
        a set of upgrades to the
        existing LIGO facilities, including frequency-dependent squeezed light,
        improved mirror coatings and potentially increased laser beam sizes.
        Noise amplitude spectral sensitivity would be improved by a factor of
        $\sim 2.5\mbox{-}3$ over 1--4\,kHz.  A+ could begin operation as early
        as 2017--18.
    \item [LIGO Voyager (LV)~\cite{ISwhitePaper}] 
        a major upgrade to the existing LIGO facilities, including higher laser
        power, changes to materials used for suspensions and mirror substrates
        and, possibly, low temperature operation.  LV would become operational
        around 2027--28 and offer noise amplitude spectral sensitivity
        improvements of $\sim 4.5\mbox{-}5$ over 1--4\,kHz.
    \item [LIGO Cosmic Explorer (CE)~\cite{ISwhitePaper}] 
        a new LIGO facility
        rather than an upgrade, with operation envisioned to commence after
        2035, probably as part of a network with LIGO Voyager.  In its simplest
        incarnation, Cosmic Explorer would be a straightforward extrapolation of
        A+ technology to a much longer arm length of 40\,km, referred to as CE1
        which would be $\sim 14\times$ more sensitive than aLIGO over 1--4\,kHz.
        An alternative extrapolation is that of Voyager technology to the
        40\,km arm length, referred to as CE2.  CE2 is only $\sim 8\times$ more
        sensitive than aLIGO for the frequency range of interest in this study.
        For simplicity, we consider only CE1. 
    \item [Einstein Telescope (ET-D)~\cite{2010CQGra..27h4007P,2011CQGra..28i4013H}] 
        the European third-generation \gw{} detector.  In this work, we consider
        the ET-D configuration which is comprised of two individual
        inteferometers where one targets low frequency sensitivity and the other
        high frequency sensitivity.  Both interferometers will be of 10\,km arm
        length and housed in an underground facility.  Furthermore, the full
        observatory will consist of three such detectors in a triangle
        arrangement.  ET-D is $\sim 8\times$ more sensitive than aLIGO over
        1--4\,kHz.  Due to the network configuration (i.e., the alignment of the
        component instruments) the effective sensitivity of ET-D is $\sim 18$\%
        higher than that for a single ET-D detector.
\end{description}
\begin{figure}
    \centering
    \includegraphics[width=0.48\columnwidth]{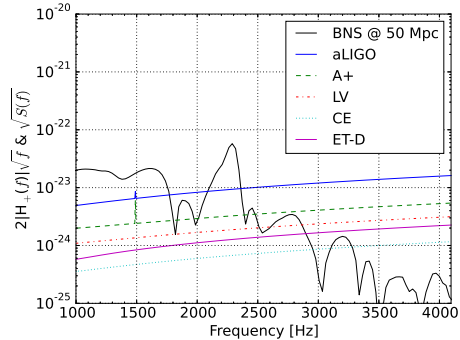}
    \caption{\label{fig:noise_spectra} Design sensitivity spectra for the \gw{}
        instruments discussed in section~\ref{sec:detectability}.  For
        comparison we also show the amplitude spectrum of a representative
        merger waveform (the TM1 EoS) evaluated at a distance of 50\,Mpc and
    optimal orientation. }
\end{figure}
Figure~\ref{fig:noise_spectra} shows the design sensitivity spectra for each
of these instruments, again focussing on the 1--4\,kHz range of interest for the
post-merger \bns{} \gw{} signal.  For comparison, we also show the amplitude
spectrum of a typical \bns{} waveform (the TM1 1.35+1.35 example discussed in
section~\ref{sec:properties}) for an optimally oriented source at 50\,Mpc.
Finally, the figures of merit describing the detectability of the post-merger
signal for each instrument are summarised in table~\ref{table:detectability}.
Note that, we compute two measures of \snr{}: SNR$_{\mathrm{full}}$, where we
simply evaluate equation~\ref{eq:inner_product} over 1--4\,kHz for the full
merger waveform; as well as SNR$_{\mathrm{post}}$, where the time-domain
waveform has been windowed to suppress power prior to the merger (taken to occur
at the peak strain amplitude), in order to yield an estimate of the contribution
to the \snr{} from the post-merger oscillations. Since there are 5 instruments,
50 waveforms and 4 figures of merit, we choose to summarise the results for each
instrument in terms of the 10$^{\mathrm{th}}$, 50$^{\mathrm{th}}$ and
90$^{\mathrm{th}}$ percentiles, evaluated over the 50 waveforms used in the
study.

\begin{table}[h]
    \centering
    \begin{tabular}{lllll}
        \toprule 
        Instrument & SNR$_{\mathrm{full}}$ &
        SNR$_{\mathrm{post}}$ & $D_{\rm hor}$ [Mpc] & $\detrate$  [year$^{-1}$]  \\
        \hline
        \hline$_{}^{}$
        aLIGO & 2.99$_{2.37}^{3.86}$ & 1.48$_{1.13}^{1.86}$ & 29.89$_{23.76}^{38.57}$ & 0.01$_{0.01}^{0.03}$ \\
        A+ & 7.89$_{6.25}^{10.16}$ & 4.19$_{3.26}^{5.35}$ & 78.89$_{62.52}^{101.67}$ &0.13$_{0.10}^{0.20 }$  \\
        LV &  14.06$_{11.16}^{18.13}$ & 7.28$_{5.64}^{9.30}$ & 140.56$_{111.60}^{181.29}$  &0.41$_{0.21}^{0.88}$  \\
        ET-D & 26.65$_{20.81}^{34.28}$ & 12.16$_{9.34}^{15.31}$ & 266.52$_{208.06}^{342.80}$ & 2.81$_{1.33}^{5.98}$  \\
        CE  & 41.50$_{32.99}^{53.52}$ & 20.52$_{15.72}^{25.83}$ & 414.62$_{329.88}^{535.221}$ & 10.59$_{5.33}^{22.78}$  \\
        \bottomrule
    \end{tabular}
    \caption{Expected detectability. The horizon distance $D_{\rm hor}$ is
        evaluated assuming an optimal matched-filter \snr{} $\rho=5$ and the
        detection rate is evaluated as described \S~\ref{sec:detectability},
        assuming the ``realistic'' binary coalescence rate
        in~\cite{2010CQGra..27q3001A}.  Large script values indicate the median
        across waveforms used in this study, while the super and subscript show
        the 10$^{\mathrm{th}}$ and 90$^{\mathrm{th}}$ percentiles, respectively.
        SNR$_{\mathrm{full}}$ refers to the \snr{} for the full waveform,
        evaluated over 1--8\,kHz.  SNR$_{\mathrm{post}}$ is the \snr{} of the
        post-merger waveform only, evaluated over the same frequency range but
        where the signal has been windowed in the time domain to suppress all
        pre-merger power.  SNRs are evaluated for an optimally oriented source
        at 50\,Mpc.  We note that the finite simulation time and numerical
        damping of the post-merger oscillations likely lead to an underestimate
        of the total \snr{} in the post-merger signal.  The reader is directed
        to the study in~\cite{2014PhRvD..90f2004C} for further
    discussion.\label{table:detectability}}
\end{table}

\section{A Waveform Model Using Principal Component Analysis}
\label{sec:pca}
The optimal data analysis method for the identification and characterisation of
a \gw{} signal in noisy data is matched-filtering, wherein an exact analytic
model, or \emph{template}, for the waveform is convolved with data stream from a
network of \gw{} detectors.   Unfortunately, the physical complexity of the
merging binary neutron star system is such that detailed numerical simulations
are required to produce even an approximate waveform.  Furthermore, since the
physical parameters of the system are essentially unknown, many such simulations
would be required in order to build a template bank to maximise the likelihood
of signal detection.

We are, therefore, confronted with a similar data analysis problem to that in
the analysis of \gw{s} from core collapse supernovae: the absence of an accurate
analytic waveform template, a limited number of computationally expensive and
approximate simulations and a requirement to significantly reduce the complexity
of the modelling problem to faciliate the use of an approximate matched-filter.
Motivated by the work in~\cite{2009CQGra..26j5005H,2009PhRvD..80j2004R}, we
find that we can construct an effective waveform model from a basis constructed
using \pca{} of a suite of merger simulations comprised of
systems with different equations of state, masses and mass ratios.

Our goal is to reduce the complexity of the modelling problem from a
high-dimensional physical parameter space, where the waveforms are modelled
directly through numerical simulation, to a lower-dimensional problem to model
the dominant features of the waveform.  \pca{} of a
catalogue of simulated waveforms provides a solution to precisely this problem.
Denoting the time-domain merger waveform as $h(t)$, its complex Fourier spectrum
is given by,
\begin{equation}\label{eq:fdomain_wave}
    \tilde{h}(\omega) = \frac{1}{2\pi} \int_{\infty} h(t) e^{i\omega t}~\diff t =
    A(f)\exp\left[i\phi(f)\right],
\end{equation}
where $A(\omega)= |\tilde{h}(\omega)|$ and
$\phi(\omega) = \arg\left[\tilde{h}(\omega)\right]$ are the magnitude and phase
spectra of signal $h(t)$, respectively,
In a similar spirit to the approach described in~\cite{2014CQGra..31s5010P} we
construct orthnormal bases for the amplitude $A(\omega)$ and phase spectra
$\phi(\omega)$ separately, using similar a \pca{}
decomposition to that described
in~\cite{2009CQGra..26j5005H,2009PhRvD..80j2004R}.  Principal component analysis
forms a basis from the eigenvectors of the covariance matrix of some set of
data.  The procedure is as follows:
\begin{enumerate}
    \item Collate a representative sample of $m$ binary merger \gw{} waveforms,
        sampled at 16384\,Hz.  This sample of waveforms is hereafter referred to
        as our \emph{training catalogue}.
        Each waveform is normalised to unit root-sum-squared amplitude
        $h_{\mathrm{rss}}$ (equation~\ref{eq:hrss}) to reduce catalogue variance
        from different amplitude scales and emphasise morphological differences.
    \item Compute the complex Fourier spectra of the time-domain waveforms in
        our catalogue. A Tukey window is initially applied to the time-domain
        signals to minimise spectral ringing and the the waveforms are
        zero-padded to a uniform $16384$ samples.  The complex spectra
        $\tilde{h}(\omega_i)$ are computed using the fast Fourier transform.
        The amplitude and phase spectra are computed from the absolute values
        and arguments of the complex frequency series and the phase spectra are
        unwrapped to yield smooth functions, each of $n=8192$ samples.
    \item \label{list:alignment} The unique feature to the analysis presented in
        this work is our choice of feature alignment, an absolutely key
        component to \pca{}.  In~\cite{2009CQGra..26j5005H}, for example, the
        \gw{} waveforms are aligned such that the peak amplitudes lie at a
        common reference time removing the need for the \pca{} to account for
        trivial variance in the catalogue.  The analogous procedure in our
        application is to align features in the frequency domain.  Each
        amplitude spectrum is rescaled such that the dominant post-merger peak,
        labelled $\omega_{\mathrm{peak}}$, is aligned to a common reference
        value $\omega_{\mathrm{align}}$.  This alignment is achieved by
        computing a set of frequencies $\omega' =
        \frac{\omega_{\mathrm{align}}}{\omega_{\mathrm{peak}}} \times \omega$,
        where $\omega$ are the angular frequencies of the original spectrum.  We
        then interpolate the original spectrum to the new frequencies where the
        dominant spectral feature (the post-merger oscillation peak) is aligned.
        Although it is not perfect, this geometric scaling (as opposed to a
        simple linear shift) also helps to align the sub-dominant $f_{2,0}$ and
        $f_{\mathrm{spiral}}$ features.  Three examples of original and aligned
        amplitude spectra are shown in figure~\ref{fig:3spec}.

\begin{figure}
        \includegraphics[width=0.48\columnwidth]{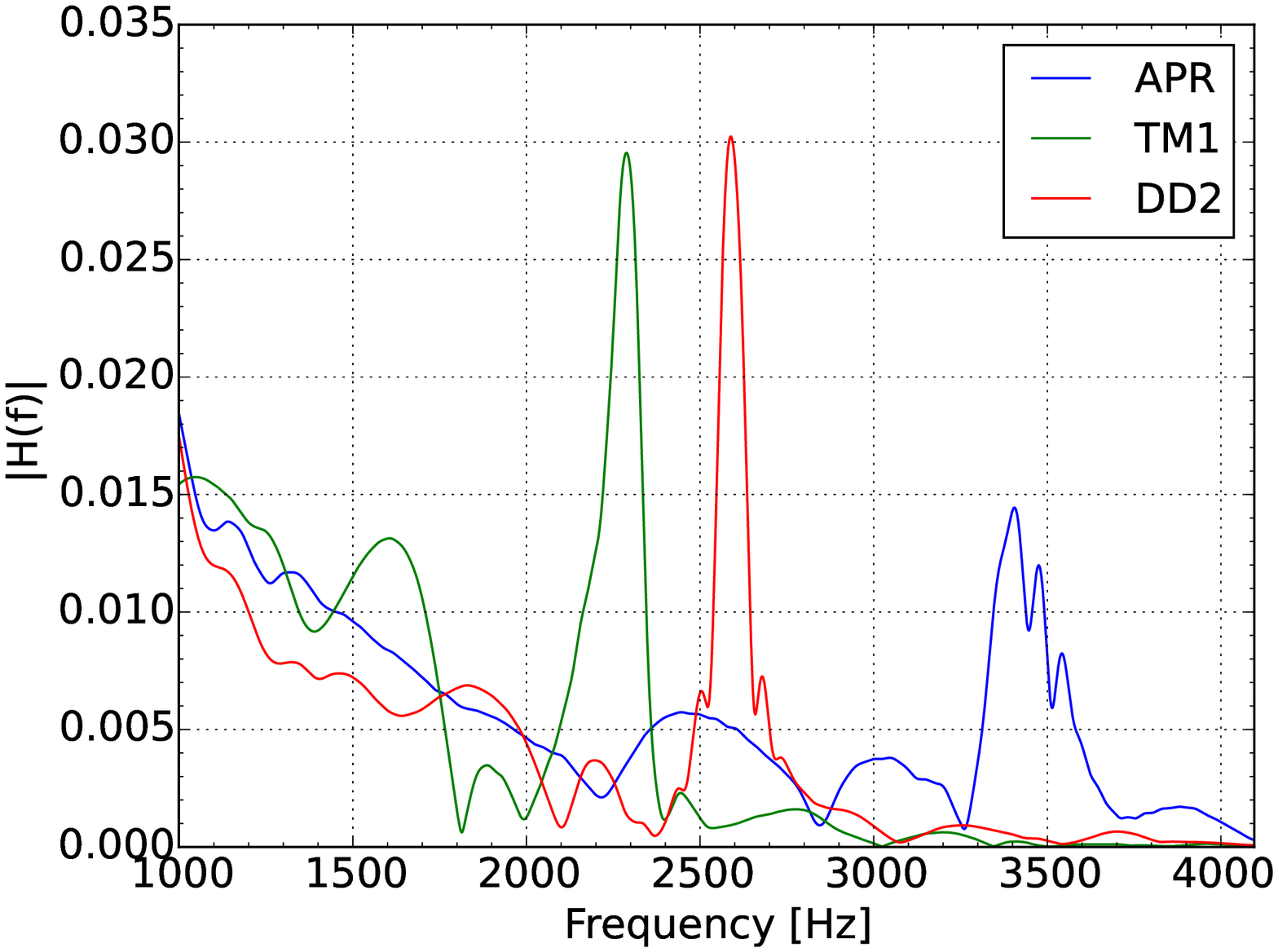}
        \includegraphics[width=0.48\columnwidth]{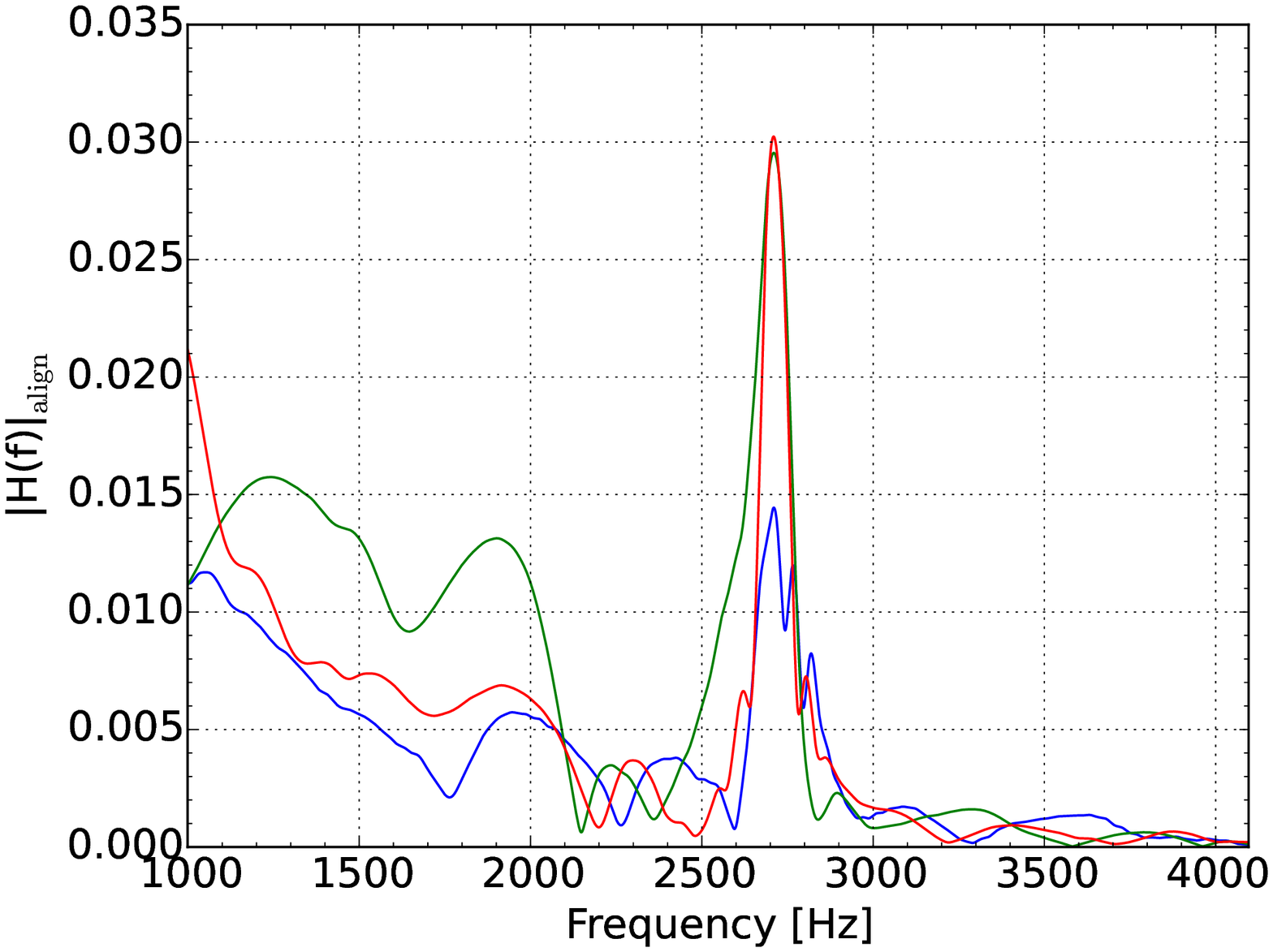}
        \caption{\label{fig:3spec}\emph{Left}: Example magnitude spectra;
        \emph{Right}: Example spectra after alignment to a common peak frequency
    $f_{\mathrm align}$.}
\end{figure}

    \item Next, we construct an $m\times n$ matrix $\mathbf{D}$ where each row
        correponds to the $n$-sample feature-aligned amplitude spectrum of each
        waveform after subtracting the mean spectrum (averaged over the $m$
        waveforms)\footnote{Note that this matrix is transposed relative to the
        descriptions in~\cite{2009CQGra..26j5005H,2009PhRvD..80j2004R} for more
    straightforward comparison with other PCA literature and use with software
packages such as that offered in~\cite{scikit-learn}.}.  The mean amplitude
spectrum, evaluated over our $m=50$ waveforms is shown in the left panel of
figure~\ref{fig:mean1pc}. 

    \item Finally, we perform the \pca{} decomposition in which we compute the
    eigenvectors of the empirical covariance matrix $\matr{D}\tran{\matr{D}}$.
    Following~\cite{2009PhRvD..80j2004R} and noting the change in row/column
    convention for the data matrix, the centered data matrix $\matr{D}$, of
    dimension
    $m\times n$, can be factorised using singular value decomposition,
    \begin{equation}\label{eq:svd}
        \matr{D} = \matr{U}\matr{S}\tran{\matr{V}},
    \end{equation}
    where $\matr{U}$ and $\matr{V}$ are orthonormal matrices with dimensions
    $m\times k$ and $n \times k$, respectively; $\matr{S}$ is a diagonal matrix
    of singular values of $\matr{D}$ in descending order and $k =
    \text{rank}(\matr{D}) \leq \text{min}(m,n)$.  The columns of $\matr{V}$,
    $\mathbf{v_1}\dots\mathbf{v_k}$, contain the eigenvectors of the covariance
    matrix $\matr{D}\tran{\matr{D}}$, our \emph{principal components}, and the
    singular values in $\matr{S}$ are the square roots of the eigenvalues
    $\lambda_i$ of the covariance matrix $\matr{D}\tran{\matr{D}}$.  Finally,
    the columns of $\matr{U}$ contain the eigenvectors of
    $\tran{\matr{D}}\matr{D}$.
    The principal components $\mathbf{v_1}\dots\mathbf{v_k}$, comprise an
    orthonomal basis of the rows (i.e., the aligned and centered waveforms) in
    $\matr{D}$ so that each of the aligned waveform amplitude (or phase) spectra
    can be represented as a linear sum of principal components and the mean.
    For example, the aligned amplitude spectrum of the first waveform can be
    constructed as:
    \begin{equation}\label{eq:reconstruction}
        A'_1(\omega) = \langle A' (\omega)\rangle + \sum_{i=1}^m \beta_i
        \mathbf{v_i}(\omega),
    \end{equation}
    where $\langle A'(\omega)\rangle$ is the mean amplitude spectrum over the
    aligned waveform catalogue and $\beta_1\dots\beta_m$ are weighting
    coefficients given by the projection of the centered $A_1(\omega)$ onto the
    principal component basis,
    \begin{equation}
        \matr{B} = \left[A'_1(\omega) - \langle A' (\omega)\rangle \right] .
        \matr{V},
    \end{equation}
    and $\beta_1,\dots,\beta_k$ are the elements of $\matr{B}$.  The original
    waveform magnitude spectrum $A_1(\omega)$ is, at last, obtained by applying
    the inverse of the alignment procedure in step~(\ref{list:alignment}).
    Figure~\ref{fig:mean1pc} shows the mean aligned magnitude spectrum $\langle
    A' (\omega)\rangle$ and the first principal component as computed for the 50
    waveforms described in section~\ref{sec:waveforms}.  Note that we have
    chosen to align the dominant post-merger peak to a value of 2710\,Hz; this
    choice is essentially arbitrary and simply corresponds to the mean of the
    peak frequencies in the catalogue.  It is important to note here that the
    value of $\fpeak$ is a free parameter in the spectral model; in practice,
    its value must be inferred from \gw{} observations.

\begin{figure}
        \includegraphics[width=0.48\columnwidth]{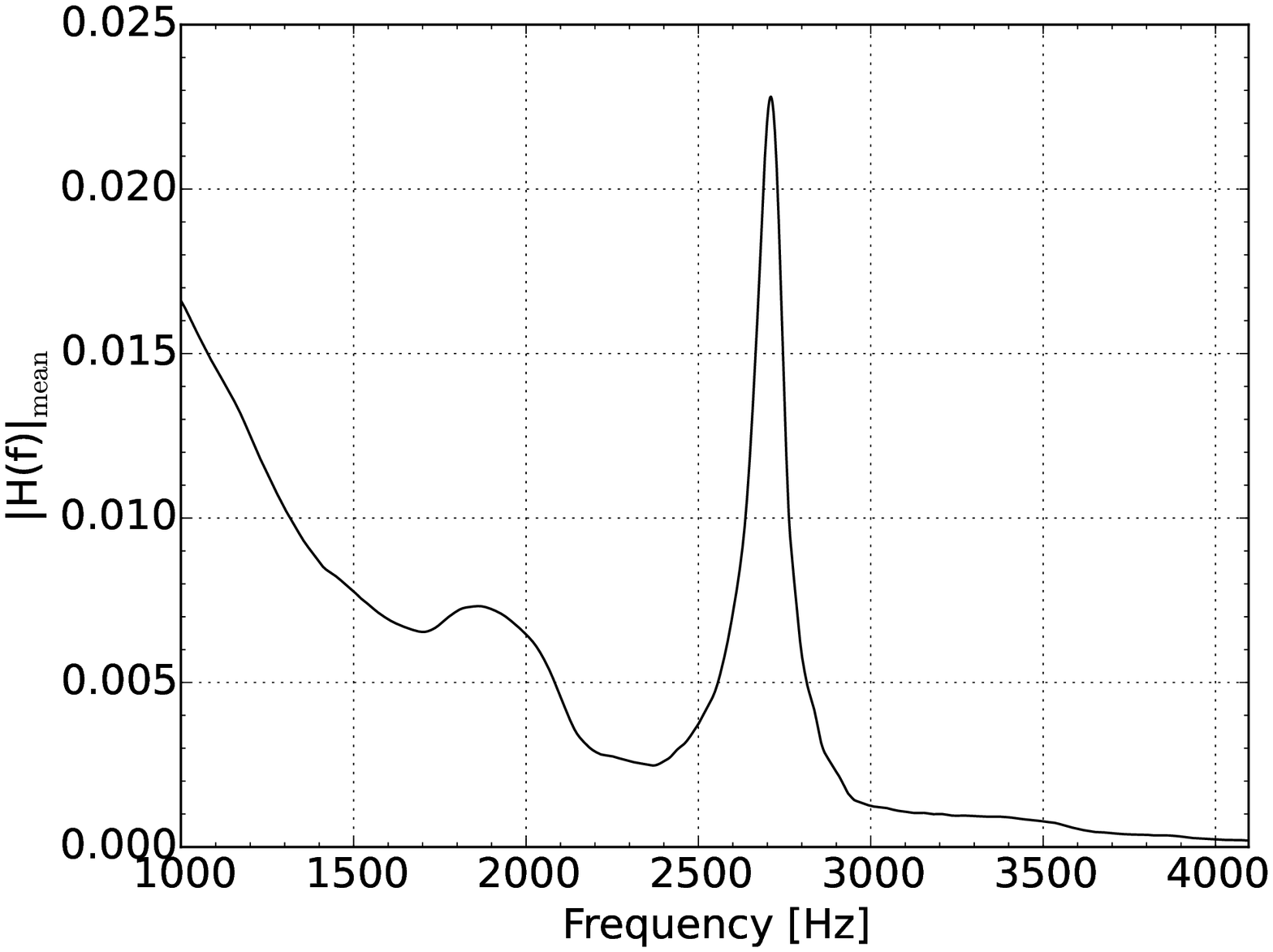}
        \includegraphics[width=0.48\columnwidth]{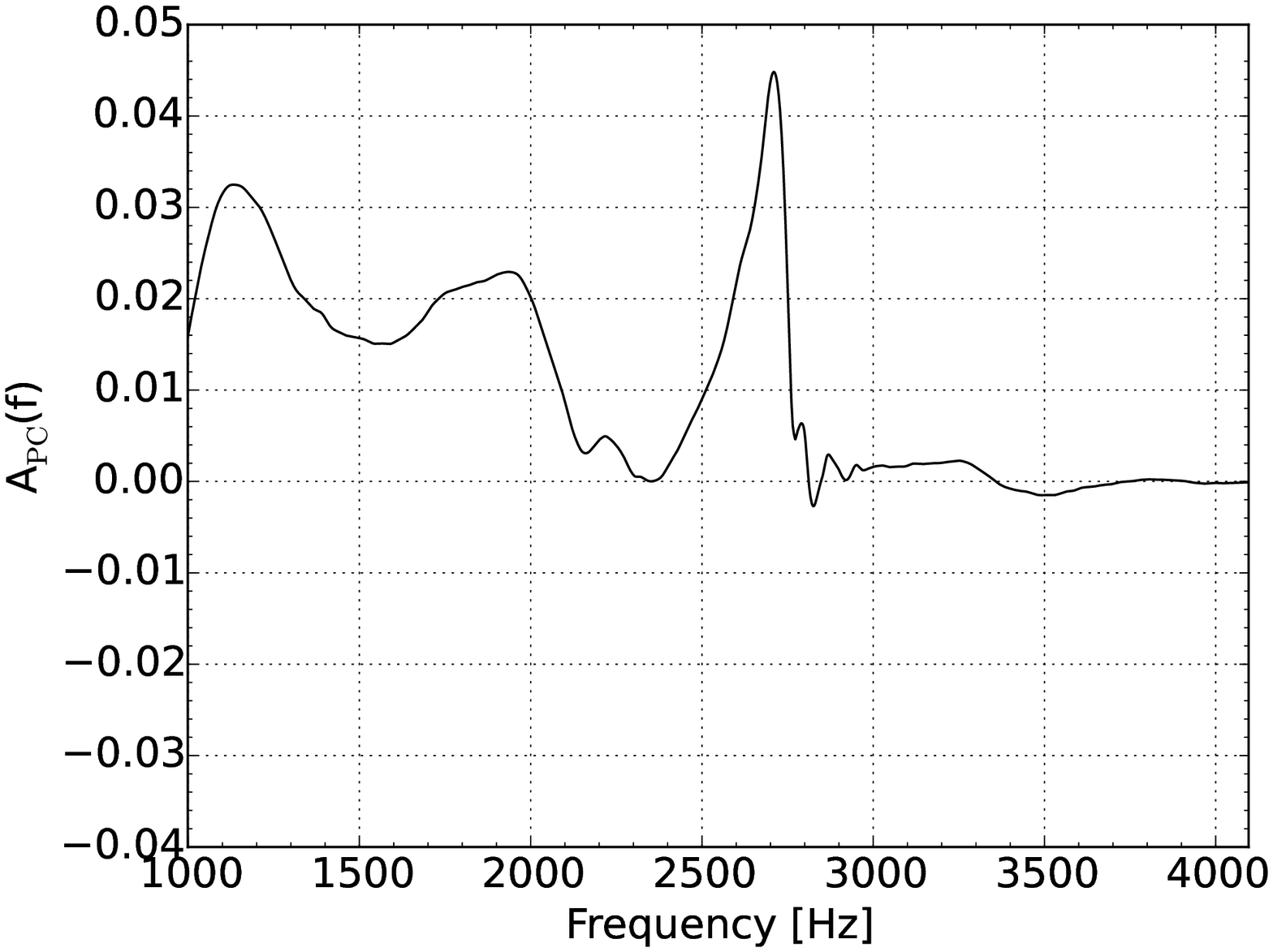}
        \caption{\label{fig:mean1pc}\emph{Left}: Mean magnitude spectrum;
        \emph{Right}: First principal component for the magnitude spectrum}
\end{figure}

\end{enumerate}
Ultimately, our goal is to construct a \emph{reduced} basis from which any
post-merger waveform can be reconstructed to some accuracy.  It is, therefore,
helpful to understand the relative importance of each principal component.  A
measure of the total variance in the centered catalogue is given by the trace of
the covariance matrix $\text{tr}(\matr{D}\tran{\matr{D}}) = \sum_{i=1}^{k}
\lambda_i$.  The variance explained by $p$ principal components is, then, the
sum of the first $p$ eigenvalues:
\begin{equation}\label{eq:explained_variance}
    \sigma^2_{\mathrm{PCA}} = \sum_{i=1}^p \sqrt{\mathbf{s}_i},
\end{equation}
\begin{figure}
    \centering
    \includegraphics[width=0.48\columnwidth]{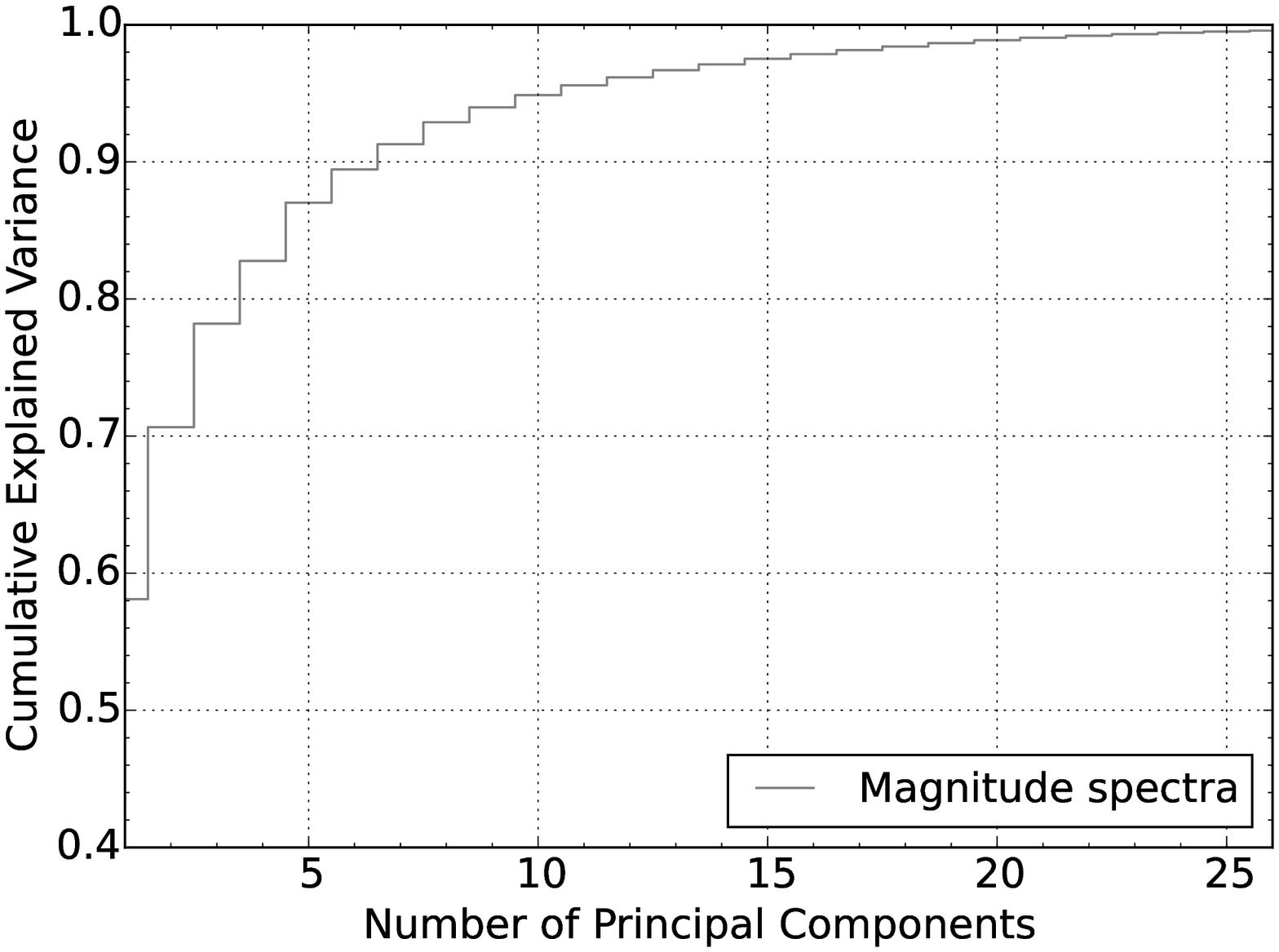}
    \includegraphics[width=0.48\columnwidth]{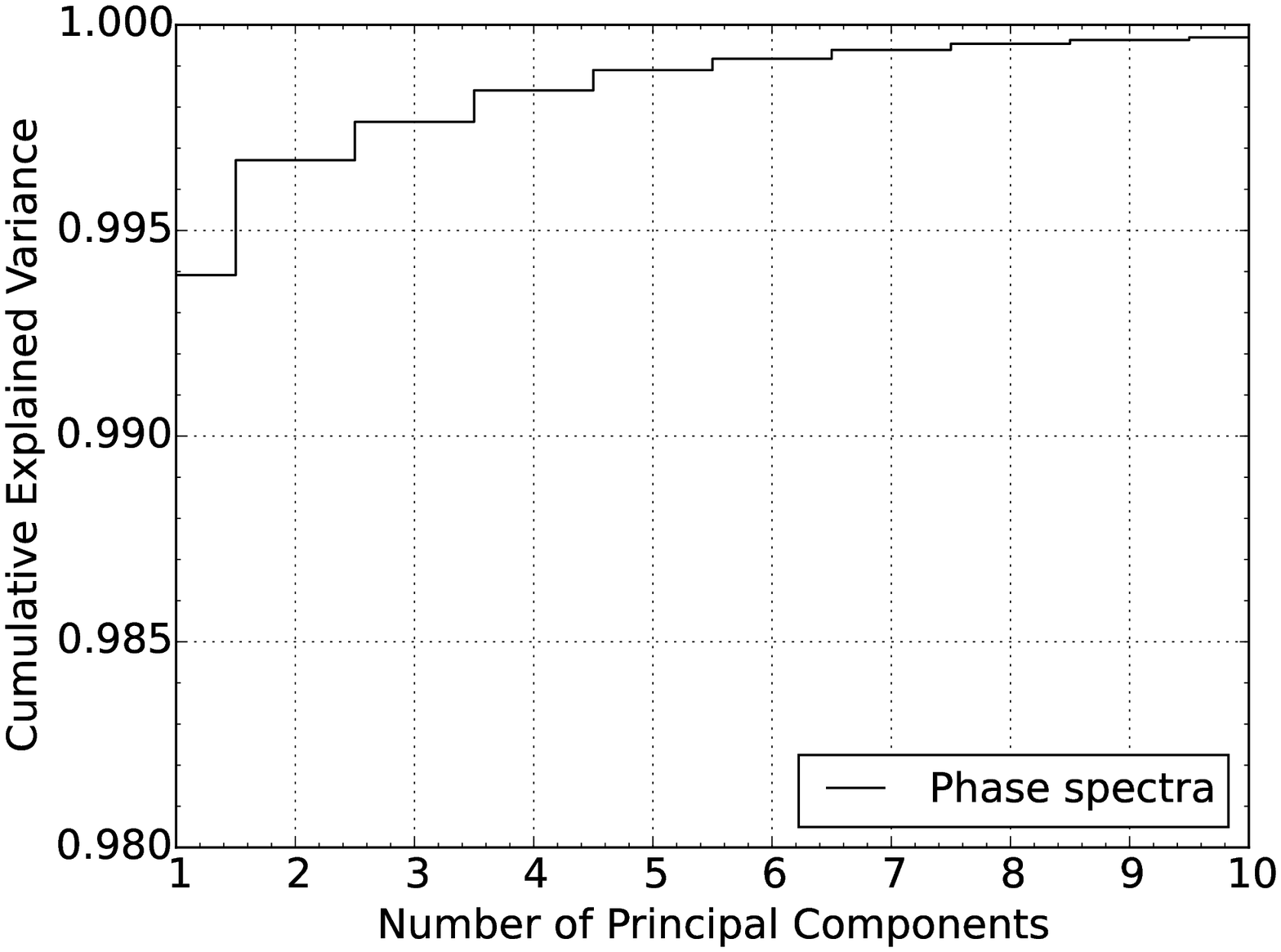}
    \caption{\label{fig:variance_lessvisc} The cumulative fraction of the
        variance within the waveform data matrix explained by each principal
        component.  \emph{Left}: results for the magnitude spectra.
        \emph{Right}: results for the phase spectra.  Note that there are 50
        waveforms in this catalogue but more than 99\% of the variance is
        explained by 25 principal components for the magnitude spectra, but only
        a single principal component is required for the phase, with most of the
    information contained in the mean phase spectrum.}
\end{figure}

\noindent where $\mathbf{s_i}$ are the singular values from equation~\ref{eq:svd}.   The
post-merger waveforms can then be approximated by using a reduced basis with
$p<m$, with the choice of $p$ based on capturing a reasonable degree of variance
in the catalogue, and equation~\ref{eq:reconstruction}.
Figure~\ref{fig:variance_lessvisc} shows the cumulative explained variance for
both the magnitude and phase spectra of the waveforms in our catalogue (i.e.,
equation~\ref{eq:explained_variance}) as a function of the number of principal
components.  One can immediately see that the variation between the waveforms is
dominated by the rich and varied structure in the magnitude spectra; only $\sim
60\%$ of the total variance is explained by the first principal component of the
magnitude spectra, while $\sim 99\%$ of the variance in the phase spectra is
described by the first component.

\subsection{PCA Templates: Characterisation \& Expected Performance}

\begin{figure}
    \includegraphics[width=0.48\columnwidth]{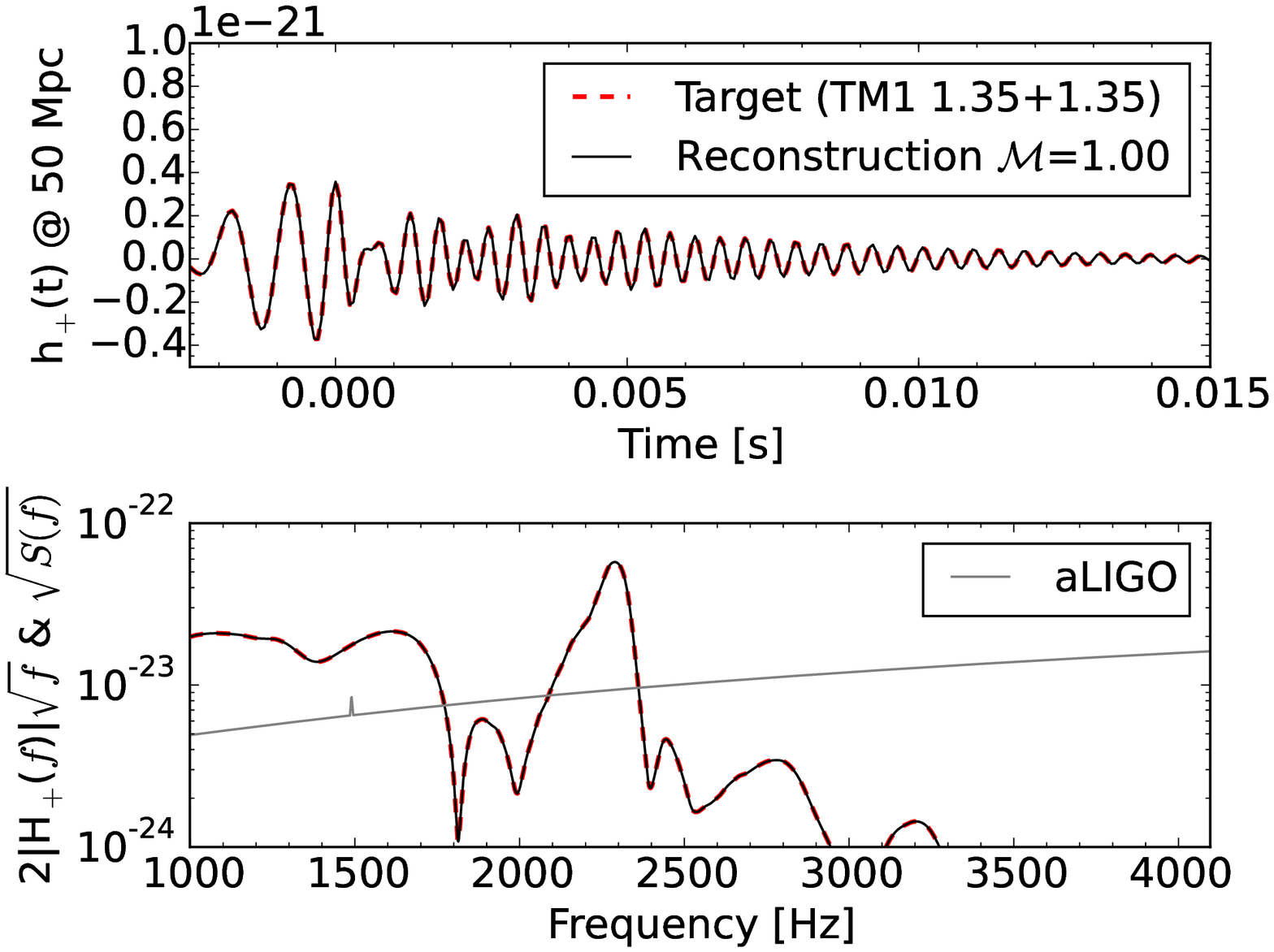}
    \includegraphics[width=0.48\columnwidth]{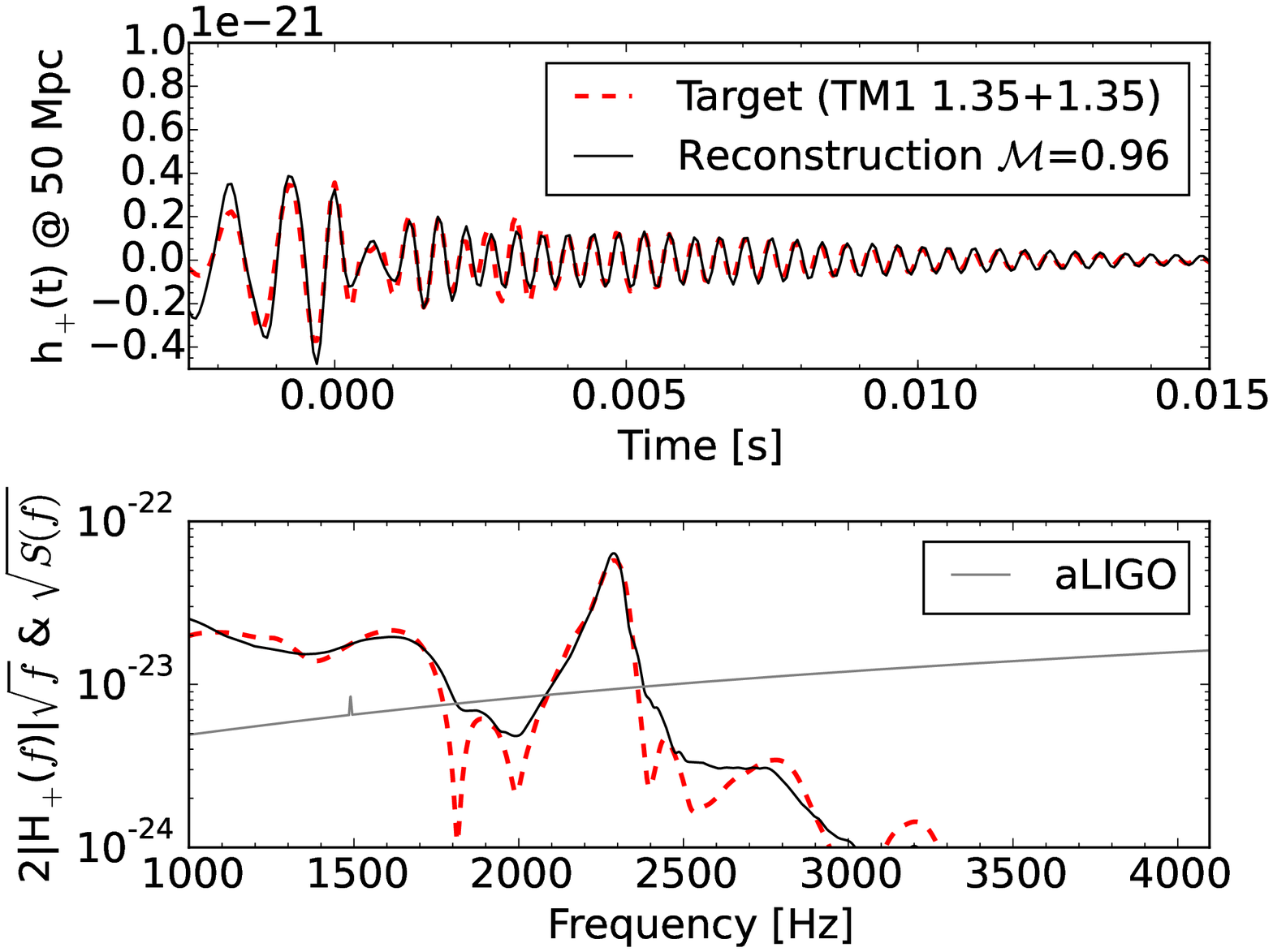}
    \caption{\emph{Left}: Reconstructing the TM1 1.35+1.35 waveform using all
    principal components. \emph{Right}: Reconstructing the TM1 1.35+1.35
waveform using only the first principal component.  The waveform has been
excluded from the waveform catalogue used to train the
decomposition. Matches in this example were computed using the aLIGO noise
curve.\label{fig:reconstruction_example}}
\end{figure}

Remembering that our goal is to build an approximate waveform template for
matched filtering, a useful figure of merit to characterise \pca{}--based model
is the waveform \emph{match} $\match$, which describes the fraction of the
optimal signal-to-noise ratio for a given signal $s(t)$ which is captured by the
waveform template $h(t)$:
\begin{equation}\label{eq:match}
    \match = \max_{t_0,~\phi_0} \frac{(s | h )}{\sqrt{(h|h)(s|s)}},
\end{equation}
where $(.|.)$ is the inner-product, defined by equation~\ref{eq:inner_product},
maximised over the start time $t_0$ and initial phase offset $\phi_0$ of the
signal.  The match is normalised such that $\match=1$ for a perfect template and
zero for an template which is orthogonal to the target signal.  In the following
examples, the match is computed assuming the aLIGO noise curve.
Figure~\ref{fig:reconstruction_example} shows an example of the reconstructed
time series and magnitude spectrum for the TM1 1.35+1.35 system considered
earlier in section~\ref{sec:properties}.  The time series is given by the
inverse Fourier transform of the complex spectrum constructed from the separate
amplitude and phase \pca{}.   As expected, when we use the full \pca{} basis
with $p=m$ and include the waveform in the data matrix $\matr{D}$, we obtain a
complete basis which allows a perfect reconstruction such that $\match=1$.

It is unlikely that nature will provide us with a signal which exactly matches
one of those contained in the set of training data $\matr{D}$.  The right panel
of figure~\ref{fig:reconstruction_example} again shows the original and
reconstructed TM1 1.35+1.35 waveform, except now this waveform has been excluded
from the training data.  In addition, we use only the first principal components
in amplitude and phase.  With this more realistic example and a much smaller
parameter space, we are still able to reconstruct the target signal with a match
$\match = 0.96$\footnote{Again, the value of $\fpeak$ is assumed known here; the
match here represents the the best case scenario}.

\begin{figure}
    \centering
    \includegraphics[width=0.48\columnwidth]{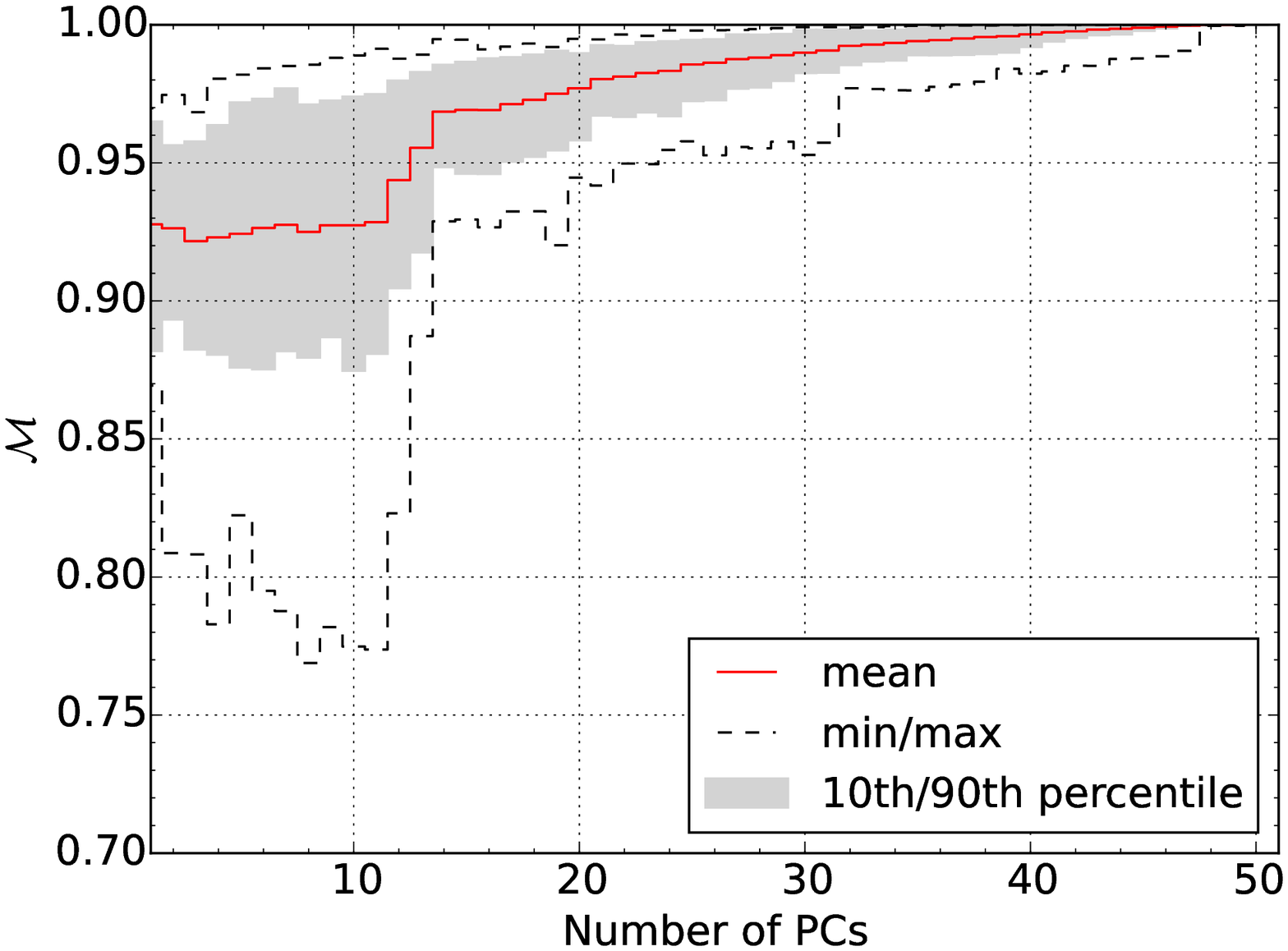}
    \includegraphics[width=0.48\columnwidth]{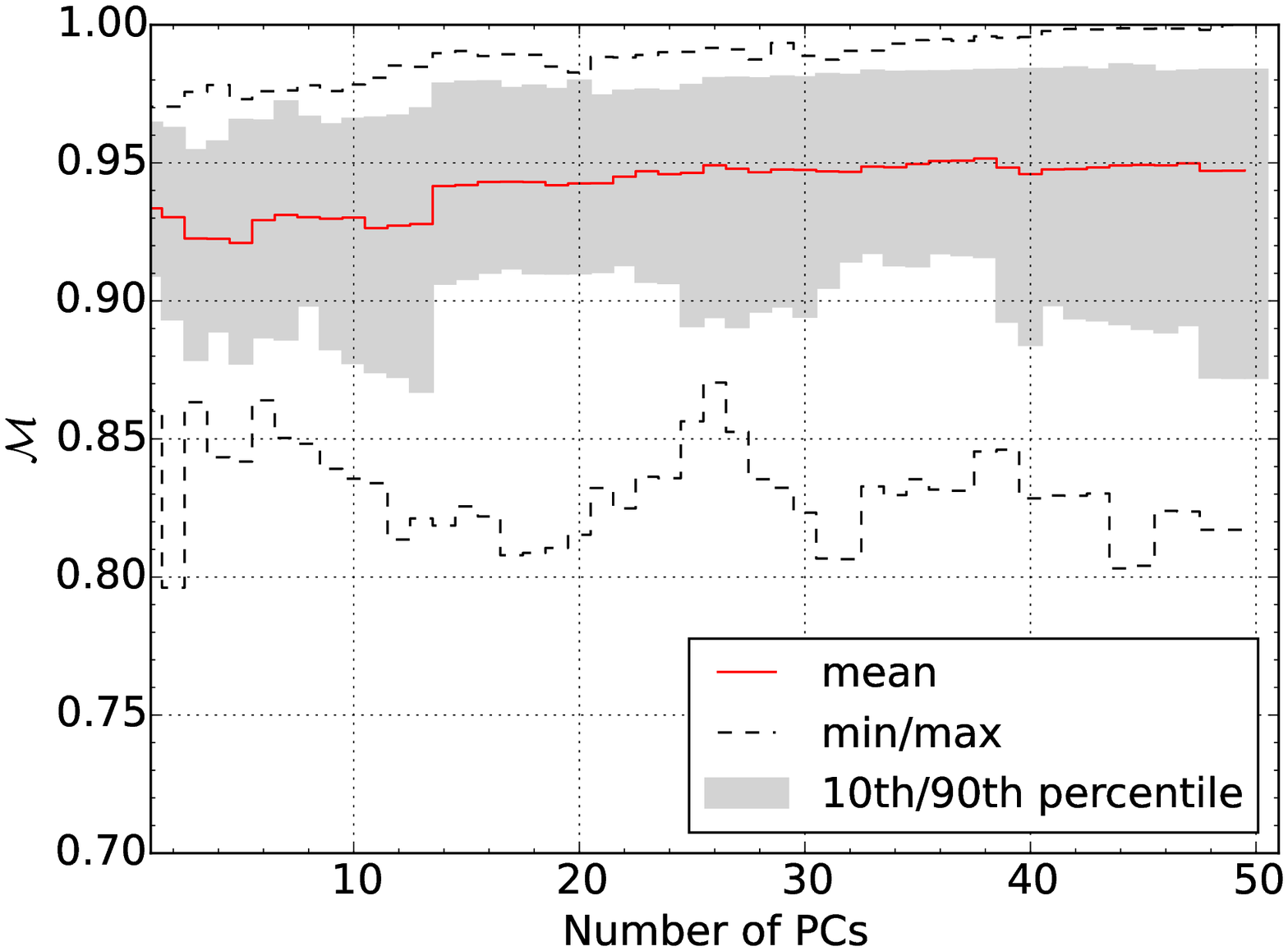}
    \caption{\label{fig:matches_lessvisc} Reconstructed waveform matches (see
    equation~\ref{eq:match}) as a function of the number of principal components
used in the reconstruction. \emph{Left}: Match when the test waveform is
included in the training data.  As expected using the full principal component
basis allows for perfect reconstruction fidelity ($\mathcal{M}=1$).
\emph{Right}: Here, matches are computed from a principal component basis
wherein the waveform whose match is calculated is withheld from the training
data.}
\end{figure}

We now compute similar matches for all of the waveforms in our catalogue and for
the different instruments described in \S~\ref{sec:detectability}.  To begin, we
compute match using the aLIGO noise curve as a function of the number of
principal components used and compare the results of including and excluding
each waveform from the data used to compute the \pca{}.  These results are
summarised in figure~\ref{fig:matches_lessvisc} with the mean, minimum, maximum
and the tenth and ninetieth percentiles over the matches computed for each of
the fifty waveforms.  The left panel shows the results when all of the waveforms
are used while the right panel summarises the matches when each waveform is
removed from the catalogue prior to computing the \pca{}.  We see that, as
before, perfect reconstruction fidelity is attained using the full basis when
all waveforms are used.  In contrast, the match remains approximately constant
with respect to the number of principal components used when each waveform being
matched is excluded from the training data.  This is a reflection of the fact
that the lower-order principal components represent the most common generic
features in the catalogue, while the higher-order components are essentially
minor corrections to the mean which may not be present in the waveform which is
excluded.  Given the quite respectable matches obtained with just the first
principal component and its apparent robustness, we propose modelling the
high-frequency \gw{} spectrum for binary neutron star mergers using
equation~\ref{eq:reconstruction} with $m=1$.

\begin{figure}
    \centering
    \includegraphics[width=0.48\columnwidth]{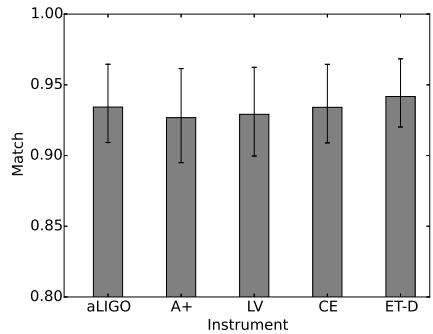}
    \caption{Template matches for each instrument discussed in
        section~\ref{sec:detectability}.  Bar height indicates the median match
        evaluated across all waveforms and error bars indicate the
        10$^{\mathrm{th}}$ and 90$^{\mathrm{th}}$ percentiles in the match.
        Templates consist of the mean waveform and a single principal component.
    When evaluating the match for each waveform, that waveform is removed from
the catalogue used for the \pca{}.\label{fig:match_summary}}
\end{figure}

We now repeat the match calculation for each of the instrument noise curves
described in~\S~\ref{sec:detectability}, using just the first principal component.
The 10$^{\mathrm{th}}$, 50$^{\mathrm{th}}$ (i.e., the median) and
90$^{\mathrm{th}}$ percentiles, computed over the matches for different
waveforms, are summarised in figure~\ref{fig:match_summary} and listed
explicitly in table~\ref{table:template_characterisation}.  We find that the
\pca{} templates yield a match of $\sim 0.93$ across all of the instruments
considered.  Variations in the match arise from differences in the shapes of the
noise curve, i.e., the denominator in equation~\ref{eq:inner_product}; in the
kHz regime, where sensitivity is limited by photon shot-noise, the noise curves
mostly only differ in their overall amplitude scale and we do not expect
significant variations in match quality.

\subsection{Implications For Parameter Estimation}
Given this approximate waveform template it is useful to determine its
effectiveness in parameter estimation and, ultimately, the extraction of
astrophysics.  Recall from~\S~\ref{sec:properties} that the single most robust
feature of the \gw{} spectrum for these signals is the presence of a dominant
spectral peak due to excitation of the post-merger remnant's quadrupolar
$f$-mode oscillation.  Recall also that the peak frequency of this excitation in
systems with total binary masses of 2.7~$M_\odot$ correlates strongly with the
radius $R_{1.6}$ of a fiducial non-spinning $1.6\msun$ \ns{} across a wide
variety of equations of state.  Our goal then, is to determine how accurately we
might expect to measure $f_{\mathrm{peak}}$ using our \pca{}--based waveform
model.  In this section, we review the $\pca{}$ waveform template and derive
Fisher-matrix estimates for the accuracy with which we may measure
$f_{\mathrm{peak}}$ and hence $R_{1.6}$ given current and planned \gw{}
observatories.

The aligned magnitude spectrum at frequency $f$ of our waveform model is given
by
\begin{equation}\label{eq:template}
        A'(f) = \langle A'(f)\rangle + \beta_1 \mathbf{v_1}(f),
\end{equation}
where the spectrum is aligned such that the dominant peak lies at frequency
$f_{\mathrm{align}}$, $\langle A'(f)\rangle$ is the mean magnitude spectrum of
the catalogue, $\mathbf{v_1}(f)$ is the first principal component and $\beta_1$
is the coefficient an arbitrary waveform's projection onto $\mathbf{v_1}(f)$.
The final magnitude spectrum is given by interpolating the aligned spectrum
$A'(f)$ to a set of new frequencies $f' =
\frac{f_{\mathrm{peak}}}{f_{\mathrm{align}}}$:
\begin{equation}
    A'(f) \rightarrow A(f) : f \rightarrow
    \frac{f_{\mathrm{peak}}}{f_{\mathrm{align}}} \times f
\end{equation}
and an identical procedure is applied to the phase spectrum $\phi(f)$.  In this
prescription then, the peak frequency $f_{\mathrm{peak}}$ is a direct parameter
of the model.

We estimate the expected accuracy of the
$f_{\mathrm{peak}}$ estimation from the Fisher matrix and by considering a
one-parameter family of waveform templates in which only $f_{\mathrm{peak}}$
varies.  We assume for simplicity that other important parameters, such as the
start time of the signal, have already been determined we hold the value of
$\beta_1$ at its nominal value.  While $\beta_1$ does indeed play a role in
determining the detailed shape of the spectrum and, particularly, the degree of
asymmetry in the main spectral peak, its will not have a strong correlation with
the location of the maximum.  A more detailed and realistic Bayesian analysis
will be conducted in the near future to account for possible correlations.

The expected error in some parameter $\theta^A$ can be determined from the
Fisher matrix $\Gamma_{AB} = (\partial_A h | \partial_B
h)$~\cite{1994PhRvD..49.2658C}:
\begin{equation}\label{eq:general_expected_error}
    (\delta \theta^A)^2 = (\Gamma^{-1})^{AA}.
\end{equation}
Following the procedure in~\cite{2009PhRvD..79l4033R}, we estimate the error in
$\fpeak$ to first order from,
\begin{equation}\label{eq:error}
    (\delta \fpeak)^2 \approx \frac{(\fpeak^{(2)}-\fpeak^{(1)})^2}{(h_2 - h_1 |
    h_2 - h_1)},
\end{equation}
where $h_1$ and $h_2$ are our waveform templates evaluated at peak frequencies
$\fpeak^{(1)}$ and $\fpeak^{(2)}$, which lie below and above the true $\fpeak$,
respectively and $|\fpeak - \fpeak^{(i)}| = \sigma_{\fpeak} \sim 1$\,Hz.   We
find that this expansion is quite stable to the choice of $\sigma_{\fpeak}$,
such that $\delta \fpeak$ varies by only a few percent up to
$\sigma_{\fpeak}\sim 5$\,Hz.

\begin{figure}
    \centering
    \includegraphics[width=0.48\columnwidth]{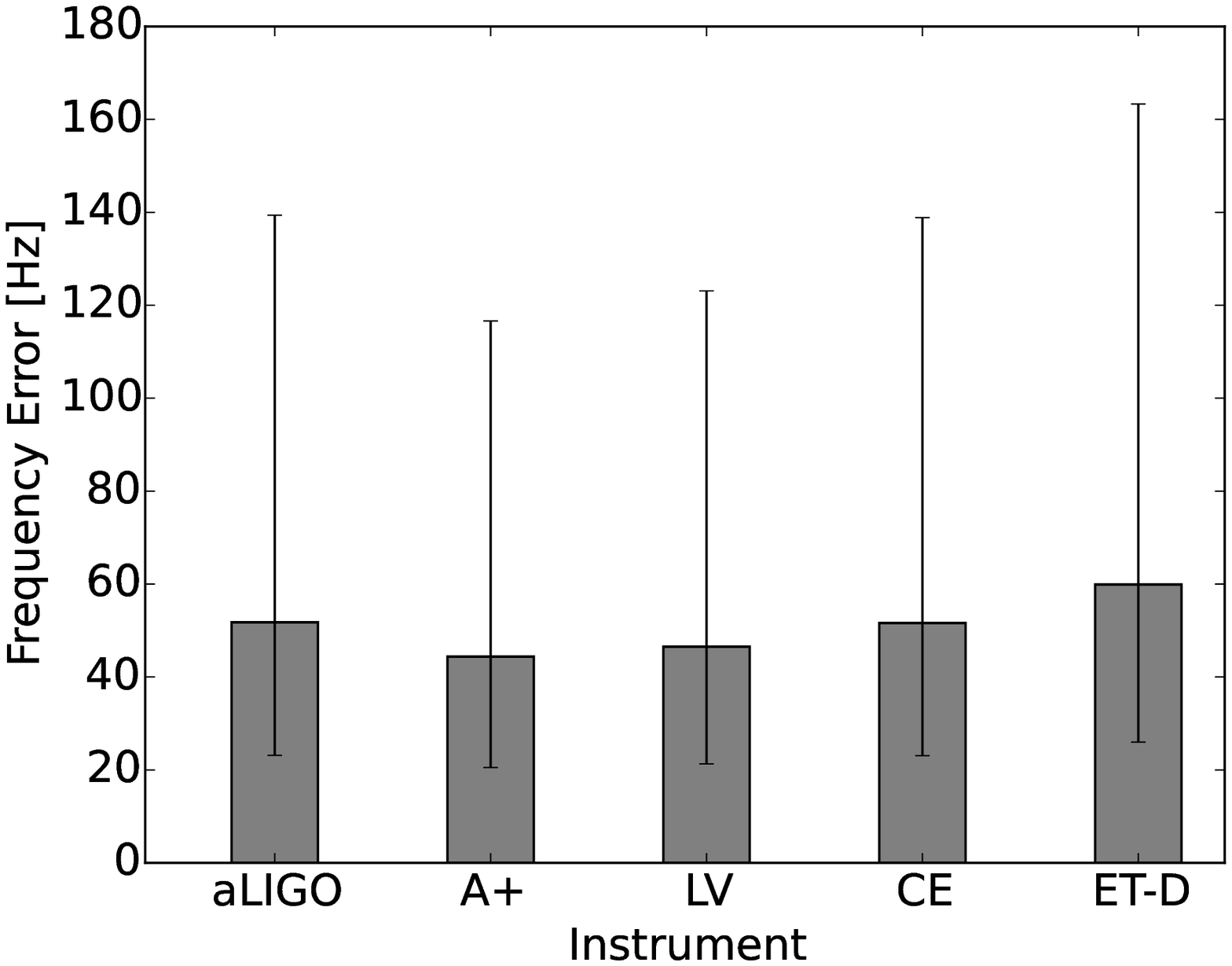}
    \includegraphics[width=0.48\columnwidth]{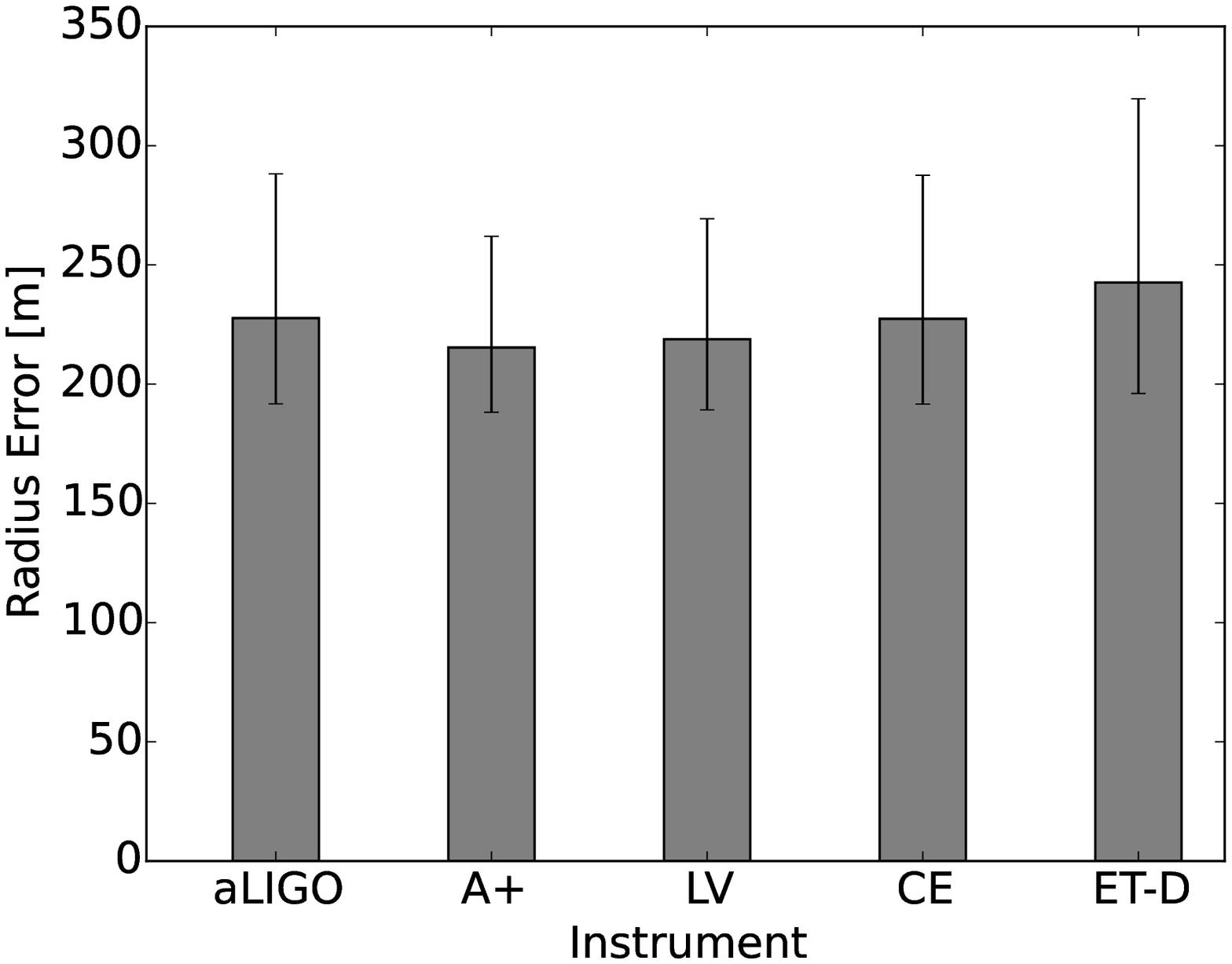}
    \caption{Expected uncertainties in parameter estimation based on Fisher
        matrix analysis of the single-PC waveform template. Bar height indicates
        the median value evaluated across waveforms and error bars indicate the
        10$^{\mathrm{th}}$ and 90$^{\mathrm{th}}$ percentiles.  \emph{Left:}
        Expected uncertainties in the determination of $\fpeak$. \emph{Right:}
        Expected uncertainties in the determination of the \ns{} radius
    $R_{1.6}$ \label{fig:pe_summary}}
\end{figure}

Figure~\ref{fig:pe_summary} and table~\ref{table:template_characterisation}
summarise the expected frequency errors obtained across waveforms and
instruments using the Fisher matrix estimate.  Errors are evaluated at \snr{}=5,
corresponding to a source at the horizon distance.  As with the match summary
from earlier the results for each instrument are summarised with the
10$^{\mathrm{th}}$, 50$^{\mathrm{th}}$ and 90$^{\mathrm{th}}$ percentiles
computed over the different waveforms.  Again, the expected frequency error is
fairly consistent between the different instruments and we find $\delta \fpeak
\approx 50$\,Hz.  We can propagate the expected error in the $\fpeak$ determination
to that in the \ns{} radius using equation~\ref{eq:freqrel}:
\begin{eqnarray}\label{eq:deltaR16}
    \delta R_{1.6}^{\mathrm{stat}} & \approx & \delta \fpeak . \frac{\partial
    R_{1.6}}{\partial \fpeak} \\
    & = & (2a\fpeak + b) . \delta f
\end{eqnarray}
The errors thus obtained represent the \emph{statistical} uncertainty in
the radius, arising from the measurement of a signal in noisy data.  The fit
given by equation~\ref{eq:freqrel} is also subject to a systematic error which,
as described in section~\ref{sec:spectra2radii}, we take to be the maximum
deviation in the $\fpeak\mbox{-}R_{1.6}$ relationship across a variety of
\eos{s}, $\delta R_{1.6}^{\mathrm{sys}}=175$\,m.  To arrive at a total expected
error in the determination of the radius $\delta R_{1.6}$ then, we quote the
quadrature sum of the statistical and systematic errors:
\begin{equation}
    \delta R_{1.6} = \sqrt{(\delta R_{1.6}^{\mathrm{stat}})^2 + (\delta
    R_{1.6}^{\mathrm{sys}})^2}.
\end{equation}
The expected radius errors thus obtained are summarised in the right panel of
figure~\ref{fig:pe_summary} and in table~\ref{table:template_characterisation}
with the usual breakdown by instrument and percentile summary statistics.

\begin{table}[h]
    \centering
    \begin{tabular}{lllll}
        \toprule 
        Instrument & $\mathcal{M}$ & $\delta f_{\mathrm{peak}}$ [Hz] & $\delta
        R_{\mathrm{1.6}}^{\mathrm{stat}}$ [m] & $\delta R_{\mathrm{1.6}}$ [m]\\
        \hline
        \hline
        aLIGO & $0.93^{0.96}_{0.91}$ & $51.8_{23.1}^{139.4}$ & $145.7_{78.3}^{228.9}$ & $227.7^{288.1}_{191.7}$\\
        A+ & $0.93_{0.89}^{0.96}$  & $44.4_{20.5}^{116.6}$  & $125.5_{69.2}^{195.0}$ & $215.3^{262.0}_{188.2}$ \\
        LV & $0.93^{0.96}_{0.90}$ & $46.5_{21.3}^{123.1}$ & $131.4_{71.9}^{204.8}$ & $218.8^{269.4}_{189.2}$ \\
        CE & $0.91^{0.96}_{0.93}$ &  $51.6_{23.1}^{138.9}$ & $78.1_{145.2}^{228.2}$ & $191.6^{287.6}_{227.4}$ \\
        ET-D & $0.94_{0.92}^{0.97}$ & $59.9_{26.0}^{163.3}$ & $168.0_{88.4}^{267.5}$ & $242.6^{319.7}_{196.1}$ \\
        \bottomrule
    \end{tabular}
    \caption{Expected template performance using the PCA methodology.
        ${\mathcal M}$ is the match given by a frequency-domain template
        composed of the mean and a single principal component, evaluated for
        magnitude and phase separately.  $\delta \fpeak$ and $\delta
        R_{1.6}^{\mathrm{stat}}$ are the expected statistical uncertainties in
        the peak frequency and \ns{} radius, respectively. $\delta R_{1.6}$ is
        the combined systematic and statistical error in the \ns{} radius,
        assuming a systematic error of
    175\,m.\label{table:template_characterisation}}
\end{table}

\section{Summary \& Outlook}
\label{sec:conclusion}
The wealth of information contained in the high-frequency spectrum of \bns{}
mergers means that there is a strong motivation to develop effective models for
the merger and post-merger phase of the coalescence \gw{} signal.  Through
consideration of the general morphology of the post-merger spectrum and the
phenomenology during and after the merger, we have determined that the
high-frequency complex spectrum is remarkably well modelled by an orthogonal
basis constructed from a catalogue of numerical simulations using a \pca{}
decomposition.

Typically, the waveform templates thus constructed yield a match of ${\mathcal
M}\sim 0.93$, over the frequency range 1--4\,kHz, with the majority of the
waveforms used in this study.  While the typical desideratum in most
matched-filtering analyses is ${\mathcal M}\gtrsim 0.97$, it is worth noting
that the only other systematic and well-quantified estimate of \gw{} search
effectiveness to date has been the burst analysis reported
in~\cite{2014PhRvD..90f2004C}.  While the burst analysis is robust to
uncertainties in the waveform it was found that its effective range was only
$\sim 40\%$ that of an optimal matched filter analysis.  The \pca{} model
presented in this work therefore holds the potential to double or even triple
the sensitivity offered by existing analyses.  Furthermore, a preliminary Fisher
matrix analysis reveals that the uncertainty in the determination of the peak
post-merger oscillation frequency is $\delta \fpeak \approx 52 $\,Hz, implying a
statistical uncertainty on the radius of a 1.6\,$\msun$ \ns{} of $\delta
R_{1.6}^{\mathrm{stat}} \approx 130$\,m for sources with sufficient power at
1--4\,kHz or proximity to Earth to produce \snr=5.  Assuming a conservative
estimate of $\delta R_{1.6}^{\mathrm{sys}}=175\,m$ for the systematic error in
the $\fpeak\mbox{-}R_{1.6}$ relation when the binary masses are known, we find
the total error in the radius is $\delta R_{1.6} \approx 220$\,m.  For
comparison, the analysis in~\cite{2014PhRvD..90f2004C} found $\delta R_{1.6}
\approx 100$\,m.  Both cases assume an $\fpeak-R_{1.6}$ relationship appropriate
for a symmetric mass configuration with total mass $2.7\msun$. Note that a) the
estimate in~\cite{2014PhRvD..90f2004C} included only the \emph{statistical}
error and b) the burst analysis requires a relatively large \snr{} before
sufficient \gw{} signal power is acquired to generate a detection candidate; by
this time, the peak frequency itself can be quite easily resolved.  Furthermore,
our estimates here are based on a single-detector analysis; those
in~\cite{2014PhRvD..90f2004C} considered a three-detector network operating with
comparable sensitivity in each instrument.

For aLIGO the horizon distance with an optimal template for \snr{}=5 is
generally $D_{\mathrm{hor}} \approx 30$\,Mpc with a plausible signal rate of
approximately 1 event per 100 years, comparable to the rate of Galactic
supernovae (see Table~\ref{table:detectability}).  Our template, however, will
lose $\sim10$\% of this \snr{}, resulting in a proportional decrease in the
horizon distance.  In fact, thanks to the local over-density of galaxies, the
impact on signal rate from this mismatch is rather negligible ($\leq 10$\%)
until we consider the ET-D or CE sensitivities.

In addition to the obvious benefit of potentially yielding a greater detection
horizon than an unmodelled search, it is worth highlighting a number of other
advantages of using even a rather ad hoc waveform template such as ours.  The
strain sensitivity spectrum of \gw{} detectors and the short duration of
post-merger \gw{} signals suggest that the pre-merger inspiral signal will
always be observed at high \snr{} for any source which is sufficiently close to
observe the post-merger signal.  This will lead to a quite precise determination
of the time of coalescence, potentially a constraint on the sky-location and
constraints on the binary masses.  If we also assume that the merger does indeed
result in the formation of a stable or quasi-stable \ns{} remnant then the
analysis need only consist of inferring the parameters of our waveform model and
does not necessarily require a signal to produce an \snr{} above some detection
threshold.  Instead, our estimate of the parameters (e.g., $\fpeak$) simply have
greater uncertainty for low \snr{} signals.  This is in contrast to typical
burst analyses which require signals to be sufficiently loud that they produce
statistically significant loud pixels in the time-frequency plane.  We note,
however, that a time-frequency \pca{} could very easily be used to `inform'
burst clustering algorithms to better target signals such as these where there
is a rich time-frequency structure.

For the purposes of this study, we adopted an \snr{} threshold $\rho_*=5$ as a
fiducial point of reference; in practice, however, it may be possible to
determine $\fpeak$ at larger distances (although with correspondingly greater
uncertainty) than suggested in this work.  Since the Fisher matrix approximation
is not valid in the low \snr{} regime, this point will be investigated via a
full Bayesian analysis using our \pca{} templates in a future study.

Finally, it is worth mentioning that the construction of \pca{} templates leads
to an intriguing and natural way to feed \gw{} observations back to the
numerical modelling community to produce a feedback loop for the estimation of
the \ns{} \eos{} and refinement of our waveform models.  For now we simply
sketch the basic algorithm as follows, leaving an example implementation to
future work:
\begin{enumerate}
    \item Construct a \pca{} template from a coarsely-sampled catalogue of
        merger waveforms, whose $\fpeak$ span the full frequency space permitted
        by allowed \eos{s} and masses.
    \item Following a nearby \bns{} detection, determine the probable component
        masses from the inspiral signal and the best estimate of $\fpeak$ from
        the \pca{} template constructed in (i).
    \item Produce a refined, more finely-sampled catalogue of merger waveforms
        (with new simulations if necessary) which correspond only to those
        \eos{s} and mass configurations which are compatible with the
        observations in (ii).  Construct a new \pca{} template from this
        catalogue.
\end{enumerate}
This process could then be iterated until some desirable stopping criterion,
such as reaching some critical value of the minimum match between the \pca{}
templates and waveforms used, is reached.  This approach may provide an avenue
to go beyond simply determining $\fpeak$ and allow an accurate reconstruction of
the full spectrum of the underlying signal.  We see then that the application of
\pca{} to construct robust and simple phenomenological templates for the
characterisation of post-merger \bns{} signals holds great promise on its own,
and may provide a useful tool to augment other approaches such as those
in~\cite{2014PhRvD..90f2004C,2015arXiv150401764B,bauswein:july15}.


\section*{Acknowledgements}
The authors thank Tjonnie Li for helpful input and careful reading of this
manuscript.  J.~C. and D.~S. gratefully acknowledge support from NSF grants
PHY-0955825, PHY-1212433, PHY-1333360, PHY-1505824 and PHY-1505524.  A.~B. is a
Marie Curie Intra-European Fellow within the 7th European Community Framework
Programme (IEF 331873). Partial support came from ``NewCompStar'', COST Action
MP1304. The computations were performed at the Max Planck Computing and Data
Facility (MPCDF), the Max Planck Institute for Astrophysics, and the Cyprus
Institute under the LinkSCEEM/Cy-Tera project.

\section*{References}
\bibliographystyle{unsrt} 
\bibliography{pmns_pca} 

\end{document}